\documentclass[options]{mn2e}
\usepackage{epsfig,amsmath}

\newif\ifAMStwofonts
\title{Chemical evolution of the M82~B fossil starburst}
\author[G.~Parmentier, R.~de Grijs, and G.~Gilmore]
       {Genevi\`eve Parmentier$^1$, Richard de Grijs$^2$, 
        and Gerry Gilmore$^2$  \\ 
    $^1$ Institute of Astrophysics and Geophysics, Universit\'e de Li\`ege,
Sart-Tilman (B5c), 4000 Li\`ege, Belgium \\
    $^2$ Institute of Astronomy, Madingley Road, Cambridge, CB3 0HA
        }
\date{Accepted .
      Received ;
      in original form }

\pagerange{\pageref{firstpage}--\pageref{lastpage}}
\pubyear{1994}

\begin{document}

\maketitle

\label{firstpage}

\begin{abstract}
M82~B is an old starburst site located in the eastern part of the M82
disc.  We derive the distributions of age and metallicity of the star
clusters located in this region of M82 by using theoretical evolutionary
population synthesis models.  Our analysis is based on the comparison of
the $BVIJ$ photometry obtained by de Grijs et al.~(2001) with the
colours of single-generation stellar populations.  We show that M82~B
went through a chemical enrichment phase up to super-solar metallicities
around the time of the last close encounter between M82 and its large
neighbour galaxy M81.  We date and confirm the event triggering the
enhanced cluster formation at about 1\,Gyr ago.  At almost the same time
an additional, distinct subpopulation of metal-poor clusters formed in
the part of M82~B nearest to the galactic centre.  The formation of
these peculiar clusters may be related to infall of circumgalactic gas
onto M82~B. 
\end{abstract}

\begin{keywords}
galaxies: individual: M82 - galaxies: starburst - galaxies: star clusters.
\end{keywords}
      
%*****************************************************
\section{Introduction}
%*****************************************************
Stellar clusters constitute unique laboratories for studying past and 
present star formation episodes in galaxies.  
Their ages and metallicities provide valuable 
insights into the star formation history as well as the chemical evolution
of the host galaxy.  Star clusters are especially important for 
understanding the history of galaxies beyond the
Local Group.  These galaxies are too far away for their 
stars to be easily resolved individually, so that most of the information
is derived from the integrated properties of star clusters, especially
of the brightest, most massive ones. 

This field has seen impressive developments during the past decade thanks to
the advent of the {\sl Hubble Space Telescope} ({\sl HST\/})
and large ground-based telescopes.
One of the main contributions to date of the refurbished 
{\sl HST} is the discovery of young clusters with typical globular 
cluster (GC) masses in galaxies beyond the Local Group.  
While the formation of 
GCs was previously thought to be restricted to the early
history of the Universe, the discovery of such candidate young GCs shows
that the formation of massive 
stellar clusters, with masses $m > 10^4 - 10^5\,M_{\odot}$, continues today. 

Young and/or intermediate age massive clusters, sometimes referred to 
as ``super star clusters'' (SSCs) by virtue of their high luminosities, 
have been detected mostly in 
interacting and merging galaxies.  The accumulation of such observations 
(e.g. Holtzman et al.~1996, Zhang, Fall \& Whitmore 2001, 
de Grijs et al.~2001) have highlighted a clear causal link 
between strong starbursts on the one hand and tidal interactions 
(e.g. the starburst galaxy M82 which is interacting with its
spiral neighbour M81) and/or mergers of galaxies (e.g. the merging pair
the ``Antennae'' NGC~4038/4039) on the other hand.

With the possible exception of the Large Magellanic Cloud, M82 is the 
nearest galaxy hosting multiple populations of such massive stellar clusters,  
thus making it a strikingly interesting object. 
Based on {\sl HST} (O'Connell et al.~1995) and ground-based imaging 
(Satyapal et al.~1997), a swarm of young stellar clusters
have been unveiled in the central regions of M82.
The galactic centre is not the only region where prolific star 
formation has recently taken place, however.  
A region extending $\sim$~400~-~1000\,pc along the major axis 
northeast of the galactic centre, labelled M82~B (nomenclature from
O'Connell \& Mangano 1978), has long been suspected 
to have also gone through a major star formation episode several hundred
million years ago.   
De Grijs et al.~(2001) obtained {\sl HST} multi-passband 
optical/near infrared observations of this fossil starburst site 
in which they discovered some 100 star clusters. \\

In this paper, we perform a detailed study of this cluster system, 
highlighting its cluster formation history and its chemical evolution.  
The paper is organized as follows.  
Section 2 briefly describes the irregular galaxy M82 and the fossil 
starburst site M82~B.  Section 3 explains how we use spectral synthesis 
models to determine simultaneously the age and the metallicity
of the M82~B clusters.  Section 4 presents the cluster formation history and 
the chemical evolution of the M82~B region as well as evidence for the
dynamical destruction over time of the clusters.  
Finally, our conclusions are presented in Sect.~5.

%*********************************************************
\section{The irregular galaxy M82}
%*********************************************************
M82 is the prototype starburst galaxy, being the nearest and, therefore, the  
best studied example of this class of galaxies.  
Freedman et al.~(1994) estimated its distance 
to be 3.6\,Mpc (i.e. its distance modulus is $m-M=27.8$).  
Such a proximity makes it uniquely valuable for a wide range 
of studies regarding extragalactic star clusters.  
M82 belongs to the nearest group of interacting galaxies, 
whose other prominent members are the large spiral M81 and a fainter 
elliptical companion, NGC~3077.  From optical images only, this set of three 
galaxies appears quiescent, showing little sign of tidal
interactions.     
However, a completely different conclusion is reached from radio surveys.  
A map of neutral hydrogen obtained with the {\sl Very Large
Array} (Yun et al.~1994) reveals huge streams of neutral hydrogen 
gas connecting the
three galaxies as well as several fainter ones.  These HI streams are dominated
by filamentary structures, loops and twists thus suggesting that the 
three largest members of the M81 group have gone through several close 
encounters.  The corresponding tidal interactions have most probably
triggered the formation of some of the M82 star clusters.  
For instance, they may
be responsible for channeling large amounts of gas into the central 
regions of the galaxy, thus inducing the active starburst in the central
regions (Rieke et al.~1993).  The study of the off-centre region M82~B
is particularly valuable since it has been shown to contain a more
evolved star cluster system, and to have the characteristics of an older 
burst of star formation (e.g., O'Connell \& Mangano 1978, Marcum \& 
O'Connell 1996, de Grijs et al. 2001). Therefore, the study of M82 B 
will enable us to probe the galaxy's history to earlier times.   \\  

Starburst regions are often obscured so heavily by dust that their study
at infrared wavelengths represents a definitive advantage with respect
to the optical.  De Grijs et al.~(2001) used the {\sl Wide 
Field Planetary Camera~2} ({\sl WFPC2}) 
as well as the {\sl Near-Infrared Camera and Multi-Object Spectrometer}
({\sl NICMOS}), on board {\sl HST}, to obtain high 
resolution imaging of M82~B in both the optical and the 
near infrared.  Two adjacent 35'' square fields, named ``B1'' and 
``B2'', were imaged, the ``B2'' region being the 
closest to the galactic centre and the active starburst.
Despite the fact that B1 and B2 simply correspond to two adjacent
fields, this differentiation, i.e. M82\,B1 and M82\,B2 instead of M82\,B,
will often be used throughout this paper as both regions show distinct 
differences in their chemical evolution (see Sect.~4.4).  
Their {\sl HST} observing run provided de Grijs et al.~(2001) 
with the $B$, $V$, $I$ and $J$ magnitudes of 
the star clusters detected in M82~B.  The availability of these four 
magnitudes allows us to disentangle the extinction, age and metallicity effects.
While the way of estimating simultaneously the age and the metallicity is 
presented in Sect.~3.1, we now briefly summarize how de Grijs et al.(2001)
derived the intrinsic colour indices of the M82~B clusters, these colour
indices being used in our study. 

In a $(B-V)$ vs $(V-I)$ diagram, the aging trajectories (i.e. the colour 
evolution with time at a given metallicity) and the extinction vector are 
not entirely degenerate for this choice of optical colors, showing that the 
$BVI$ diagram could be used to estimate the colour excesses by which a cluster
is reddened (de Grijs et al.~2001, their Fig.~12).
This method also allows to derive extinction values for each cluster individually.
This is useful in a galaxy like M82, which has a significant 
and highly variable extinction and which cannot be characterized by
a single mean value for the internal extinction and reddening.   
Assuming that the metallicity of the M82~B clusters is roughly solar, 
de Grijs et al.~(2001) used this property to estimate their intrinsic colour 
indices, their extinction and their absolute magnitudes.

Using the colour indices dereddened as explained above, 
de Grijs et al.~(2001) determined the age distribution of the
M82~B clusters, which contains information about 
the cluster formation rate.  The age of the M82~B clusters ranges from 200~Myr 
to over 10~Gyr, a time interval over which clusters have been forming 
continuously, although not at a constant rate.  
The most striking feature is a strong peak of cluster 
formation starting about 1\,Gyr ago and lasting about 600\,Myr
(de Grijs et al.~2001), a time roughly corresponding to the 
last perigalactic passage of M81 and M82 (Brouillet et al.~1991).
The close match between both of these independently determined time-scales
provides, therefore, very strong evidence that interactions between
galaxies lead to enhanced star {\it cluster} formation.

While de Grijs, Bastian \& Lamers (2002) study the dynamical 
history of the M82~B star cluster system in great detail, we focus on its 
chemical evolution.  In order to carry out such an analysis, we refine 
the results obtained by de Grijs et al. (2001), that is, we redetermine 
both the age and the {\it metallicity} of the star clusters.

%***********************************************************
\section{Spectral Synthesis}
%***********************************************************
Since the pioneering papers of Tinsley (1972) and Searle, Sargent \& Bagnuolo 
(1973), spectral population synthesis, i.e. the modeling of the 
spectral energy distribution of a star cluster, 
has become a standard technique to study the integrated stellar 
populations of galaxies.  Assuming that stellar clusters are 
single-age and single abundance groups of stars, their integrated colours 
reflect their age and metallicity (for a given initial mass function, IMF). \\

In order to perform a self-consistent comparison of our results with those
obtained by de Grijs et al.~(2001), we use the same spectral 
synthesis model, i.e. the Bruzual \& Charlot (1996, 
hereafter BC96) model.  We consider the photometric properties of 
Simple Stellar Populations (SSPs; i.e. single-age and single abundance 
groups of stars) with a Salpeter IMF from 
0.1\,M$_{\odot}$ up to 125\,M$_{\odot}$.  
BC96 provides the temporal evolution of the broad-band colours of such
SSPs for a wide range of metallicities.  In the following, we consider 
the three intrinsic colour indices $(B-V)_0$, $(V-I)_0$ and $(V-J)_0$ 
whose observational counterparts were obtained by de Grijs et al.~(2001).
We show below how the addition of the $(V-J)_0$ colour index enables 
us to estimate both the age and the metallicity of the clusters.  

%-------------------------------------------------------------------
\subsection{Disentangling age and metallicity effects} 
%-------------------------------------------------------------------
A greater age or a higher metallicity both lead to redder colours,
an effect known as the age-metallicity degeneracy.  
For instance, in a $(B-V)_0$ versus $(V-I)_0$ plot, the iso-metallicity
tracks (i.e. the locus of an SSP at different ages for a given 
metallicity) and the isochrones
are parallel and located in the same part of the optical colour-colour
diagram, precluding therefore any simultaneous estimate of age and 
metallicity.  
De Grijs et al.~(2001) also obtained the {\it J}-band magnitude for
each cluster.  Figure 1 explores the suitability of a $VIJ$ 
colour-colour diagram to lift this age-metallicity degeneracy.  
Thin lines represent isometallicity tracks, i.e. aging trajectories,
and thick lines refer to isochrones (from left to right,
log\,$t/{\rm yr}$\,=\,8.5, 9.0, 9.5 and 10.0).  
The grid defined by the isometallicity tracks
and the isochrones shows that a $(V-J)_0$ vs $(V-I)_0$ diagram is well
suited to disentangle the effects of age and metallicity.  \\

As mentioned in Sect.~2, the cluster intrinsic colours were determined 
by de Grijs et al.~(2001) based on the hypothesis that the metallicity 
of the clusters in M82~B was roughly solar.  However, the next section
shows that M82~B hosts very metal-poor (Z $\simeq$ 0.02Z$_{\odot}$)
and super-solar abundance (Z $>$ Z$_{\odot}$) clusters in addition to solar
abundance ones.  Therefore, one can question the validity of the 
present analysis which makes use of the intrinsic colours determined by de
Grijs et al.~(2001).  
The solar metallicity track and the extinction trajectories being not 
entirely degenerate in a $(B-V)_0$ vs $(V-I)_0$ diagram, 
de Grijs et al.~(2001) were able to estimate the visual extinction and the
colour excess for each cluster (see Sect.~2).  Furthermore, the 
iso-metallicity tracks are at the same locus of the diagram, with the distance 
between two subsequent tracks of distinct metallicity being 
smaller than the photometric uncertainties.  As a consequence, the visual
extinction values and the colour excesses (or, equivalently, the intrinsic 
colours) estimated by de Grijs et al.~(2001) on the basis of the solar 
metallicity hypothesis, can be used throughout this work irrespective of  
the metallicity of the clusters.   

\begin{figure}
\hspace*{-5mm}
\epsfig{figure=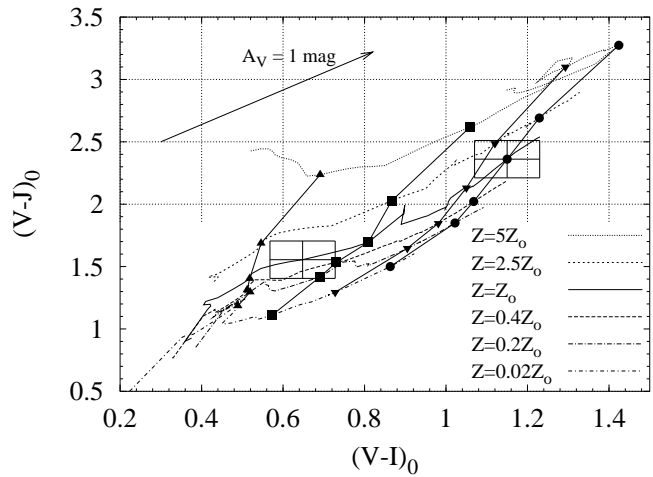, width=1.1\linewidth} 
\label{fig:VIJ}
\caption{$(V-I)_0$ vs $(V-J)_0$ diagram of SSPs based on the BC96
model.  The six thin curves represent isometallicity tracks, i.e.
the locations of SSPs of different ages for six distinct metallicities.
The four thick curves connecting filled symbols represent four isochrones
(from left to right: triangles, log~$t/{\rm yr}$\,=\,8.5; squares, 
log~$t/{\rm yr}$\,=\,9; reversed triangles, log~$t/{\rm yr}$\,=\,9.5; 
circles, log~$t/{\rm yr}$\,=\,10).  
Each filled symbol is an intersection between an isochrone 
and an isometallicity track.  The thin boxes represent the 
medians of the 1$\sigma$ error distributions of the sample in $(V-I)$ 
and $(V-J)$ for two SSPs of solar metallicity
at ages of 600\,Myr and 10\,Gyr.  The arrow indicates the effects 
of reddening on these models, for a visual extinction 
$A_V = 1\,{\rm mag}$.}  
\end{figure}

%-------------------------------------------------------------  
\subsection{Accuracy of the age and metallicity estimates 
for unextinguished clusters}
%---------------------------------------------------------------
The medians of the $1\sigma$ error distributions for the different colour 
indices are $\sigma_{(B-V)} \simeq$ 0.08, $\sigma_{(V-I)} \simeq$ 0.08 and 
$\sigma_{(V-J)} \simeq$ 0.15.  Let us first assume that the 
uncertainties of the observed colours constitute a good approximation
of the intrinsic colour uncertainties. 
The typical uncertainties in $(V-I)$ and $(V-J)$ are superimposed 
on the age-metallicity grid in Fig.~1 for the solar metallicity SSPs
of 600\,Myr and 10\,Gyr.  The corresponding 
uncertainties in age are the smallest for SSPs with ages from 
300\,Myr to 1\,Gyr.  For instance, for an SSP of solar metallicity the 
typical error in the estimated age is not greater than 40\% at an age 
of 600\,Myr, while it is as large as a factor of 2 for a 10\,Gyr
old SSP.  This greater uncertainty at greater age, which is related to the
tightening of the isochrones at log $t/{\rm yr} \ge 9.5$  
in the $VIJ$ diagram, is another aspect of 
the decrease in sensitivity of the colour indices to ages as an SSP ages 
(e.g. Worthey et al.~1994).  Interestingly, the diagram shows 
the greatest sensitivity to ages within the time 
interval of interest to us, namely the age of the burst of cluster 
formation at $\log t/{\rm yr} \sim 9$.  \\
As for the accuracy of the metallicity estimates, the differences in
the $(V-J)_0$ colour index between the isometallicity tracks 
Z\,=\,Z$_{\odot}$, Z\,=\,0.4Z$_{\odot}$ and Z\,=\,0.2Z$_{\odot}$ are 
of a similar order of magnitude as the uncertainty $\sigma_{(V-J)}$ so that
we cannot easily distinguish among such close-to-solar metallicities
based on the available data set.  
However, the $VIJ$ colour-colour diagram can, in fact, be used to distinguish 
solar metallicity SSCs from very metal-poor (i.e. Z\,=\,0.02Z$_{\odot}$) 
and super-solar abundance (i.e. Z $>$ Z$_{\odot}$) populations.  

We caution however that this reasoning does not take into account
the {\sl extinction} uncertainties.  Figure 1 displays the extinction 
vector in the $VIJ$ diagram and clearly shows that extinction uncertainties 
cannot be neglected as they will combine with the colour uncertainties quoted 
above to enlarge the error boxes in Fig.~1.  This reasoning therefore
applies only to clusters whose intrinsic and observed colours are not 
significantly different.
In Sect.~4.2, we will estimate in a more careful way how the magnitude 
uncertainties (i.e. $\sigma _B$, $\sigma _V$, $\sigma _I$, $\sigma _J$) 
propagate as extinction, age, mass and metallicity uncertainties.

%**************************************************
\section{The Cluster Formation History of M82~B}
%**************************************************
In this section, we estimate the age, the metallicity and the mass-to-light 
ratio of the M82~B clusters by comparing their extinction-corrected 
colours to those predicted as a function of age and metallicity by BC96.  
We will then derive the cluster formation history of the M82~B region.  

%------------------------------------------------------------
\subsection{Age, metallicity and mass estimates}
%------------------------------------------------------------
The age ($t_{\rm cl}$) and the metallicity ($Z_{\rm cl}$) of each stellar cluster 
are estimated simultaneously by means of a least-squares minimisation: 
\begin{equation}
{\rm min} \left\{S^2(t,Z) = \sum_{i=1}^3 \left(\frac{{{\rm CI}_i}^{\rm intr} - 
{{\rm CI}_i}^{\rm SSP}(t,Z)}{\sigma_{{\rm CI}_i}}\right)^2\right\}  
\end{equation}
\begin{flushleft}
~~~$\Rightarrow$ ~~ t$_{\rm cl}$ and Z$_{\rm cl}$~.
\end{flushleft}
In this relation, ${\rm CI}_{i=1,2,3}^{\rm intr}$ denote the three intrinsic 
colour indices (i.e. $(B-V)_0$, $(V-I)_0$, $(V-J)_0$) and 
CI$_{i=1,2,3}^{\rm SSP}(t,Z)$ represent the corresponding BC96 
integrated colours of an SSP with age $t$ and metallicity $Z$.  
$\sigma _{{\rm CI}_{i=1,2,3}}$ are the 1$\sigma$ uncertainties in the 
colour indices.  The resulting age and metallicity distributions are
shown in Figs.~2 and 3, respectively. They will be discussed, along with the
other results, in the next sections.  
The different panels of Fig.~\ref{fig:compareCI} 
display the comparison between 
the BC96 colour indices for the derived ages and metallicities
and the corresponding intrinsic colour indices 
obtained from the multicolour {\sl HST} photometry.  
The 1$\sigma$ error bars and, when relevant, the upper/lower limits 
(indicated by the arrows) of the observed colours are included.  
The good agreement between the observed colour indices and those derived  
from the BC96 spectral synthesis models illustrates the reliability 
of our age and metallicity estimates.  \\

Using these age and metallicity estimates,
we can now apply the age and metallicity dependent mass-to-light ratio
$\frac{M}{L_V}(t,Z)$
predicted by BC96 for a single burst stellar population to derive
masses $m$ for our cluster sample.

As quoted above (Sect.~3), the mass-to-light ratio predicted by 
the BC96 model we use
assumes a Salpeter IMF.  However, there is now ample evidence that
a single power-law Salpeter IMF can be ruled out at masses lower than 
1\,M$_{\odot}$, the IMF slope becoming clearly much shallower than the 
Salpeter IMF.  For instance, the IMF observed in the solar 
neighbourhood may be roughly flat at the lower end of the stellar mass 
spectrum or may even peak at a mass  around 0.25\,M$_{\odot}$ and 
decline into the brown dwarf regime (Larson 1998).  The Salpeter IMF
obviously overestimates the relative number of subsolar-mass stars.
These stars being those with the largest mass-to-light ratio, 
a Salpeter IMF causes an overestimate of the individual cluster 
masses, {\sl although the relative mass distribution of our entire
cluster sample remains unaffected}.  If we had used more modern IMFs
(e.g. Kroupa, Tout \& Gilmore 1993) this would have resulted in lower cluster mass estimates,
by a factor $\simeq$ 2-3 (see de Grijs et al.~2002) with respect to
the mass estimates presented in this paper.   \\

Our best age, mass and metallicity estimates for each cluster, along with 
their uncertainties (see the following section) are listed in Tables 1 
and 2 for B1 and B2, respectively.   The results are presented in 
Figs.~2 to 9.  In these figures, the few clusters for which either 
$B$, $V$ or $I$ is an upper or lower limit only are represented 
by open symbols or dashed histograms.

%-------------------------------------------------------------
\subsection{Uncertainty estimates}
%--------------------------------------------------------------
We now address the issue of 
estimating the errors by which the derived cluster properties are affected.  
Since the extinction was derived from the $BVI$ diagram, it is clear that
uncertainties in the optical filters will propagate as extinction 
uncertainties.  In the $BVI$ colour-colour diagram, the extinction is the
distance, measured along the reddening trajectory, from the observed point
($(B-V)$,$(V-I)$) down to the aging trajectories (Fig.~12 of de Grijs et 
al.~2001).  Considering stellar populations older than 100\,Myr, the age
range of the vast majority of the clusters of our sample, the aging 
trajectories display the same linear growth of $(V-I)_0$ vs  $(B-V)_0$, 
whatever the metallicity.  It is therefore reasonable to approximate 
all the iso-metallicity tracks by the same straight line with slope and 
ordinate origin $m_{\rm SSP}$ ($\simeq 0.87$) and $p_{\rm SSP}$ 
($\simeq$ 0.28), respectively:
\begin{equation}
(V-I)_0 \simeq m_{\rm SSP} (B-V)_0 + p_{\rm SSP}\;.
\label{SSP_BVI}
\end{equation}    
Under this assumption, it is easy to show that the extinction modulus 
obeys:
\begin{multline}
A_V \simeq \frac{\sqrt{1+m_{\rm ext}^2}}{\sqrt{E(B-V)^2+E(V-I)^2}} \\
\times \frac{(V-I)-m_{\rm SSP}(B-V)-p_{\rm SSP}}{m_{\rm ext}-m_{\rm SSP}}\;.
\label{Av}
\end{multline}    
In this equation, $E(B-V)$=0.32 and $E(V-I)$=0.52 are the optical colour 
excesses corresponding to an extinction of one magnitude, $m_{\rm ext}$ 
is the slope of the reddening trajectory, i.e. $m_{\rm ext}=0.52/0.32$
(Cardelli, Clayton \& Mathis 1989).
Substituting these numerical values into Eq.~(\ref{Av}), we get an estimate of the
extinction dimming a cluster whose measured optical magnitudes are $B$, $V$ and $I$:
\begin{equation}
A_V \simeq 4.14 \left[ (1+m_{\rm SSP})V -m_{\rm SSP}B -I -p_{\rm SSP} 
\right]\;.
\label{Av_coef}
\end{equation}    
If we assume that the $BVI$ measurements are entirely independent, the $1\sigma$ error by 
which the extinction is affected derives from Eq.~(\ref{Av_coef}):
\begin{equation}
\sigma_{A_V} \simeq 4.14 \sqrt{m_{\rm SSP}^2 \sigma _B^2 
+ (1+m_{\rm SSP})^2 \sigma _V^2 + \sigma _I^2}\;.
\label{sig_Av}
\end{equation}    
For each cluster, Tables 1 and 2 also list the best, minimal and maximal values
of the extinction $A_v$.  The observed points of a few clusters fall below 
the aging trajectories in the $BVI$ diagram but remain nevertheless consistent 
with the theoretical tracks within the photometric uncertainties (de Grijs 
et al.~2001, their Fig.~12).  In such cases, we have assumed zero extinction. \\

\begin{figure}
\begin{minipage}[b]{0.995\linewidth}
\begin{center}
\epsfig{figure=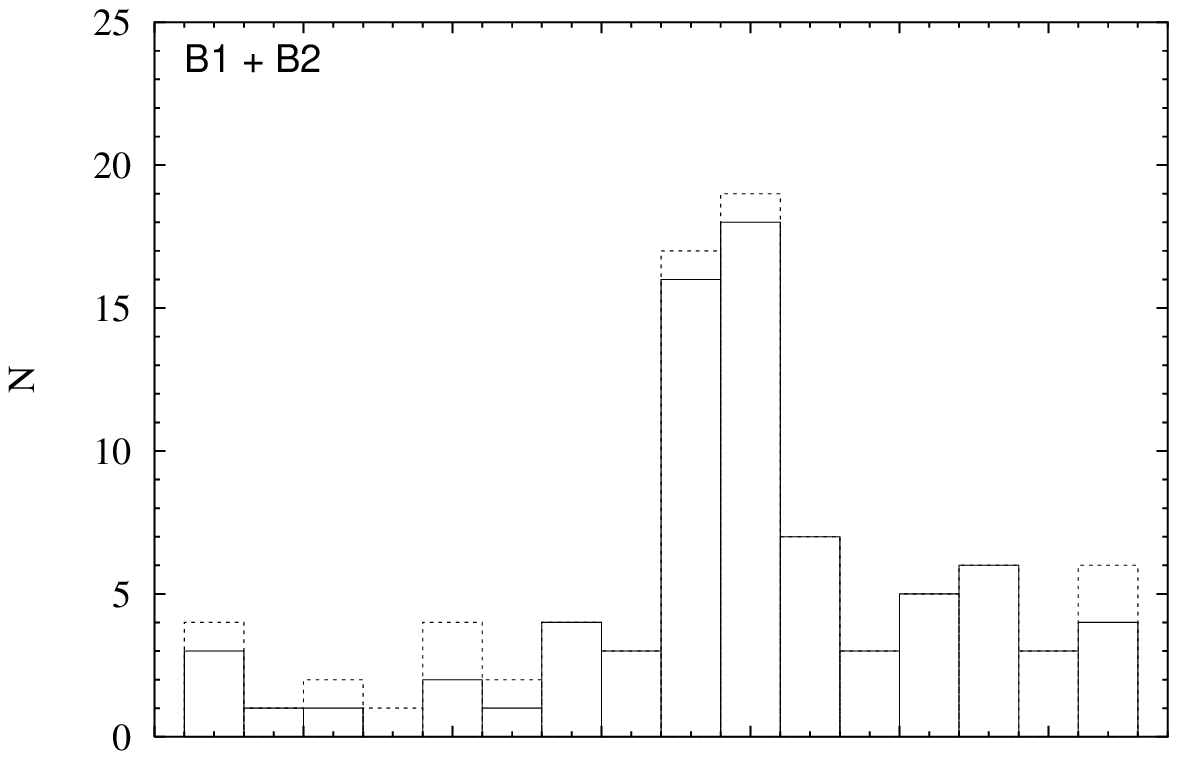, width=\linewidth}
\end{center}
\end{minipage}
\vfill
\vspace*{-9mm}
\begin{minipage}[b]{0.995\linewidth}
\begin{center}
\epsfig{figure=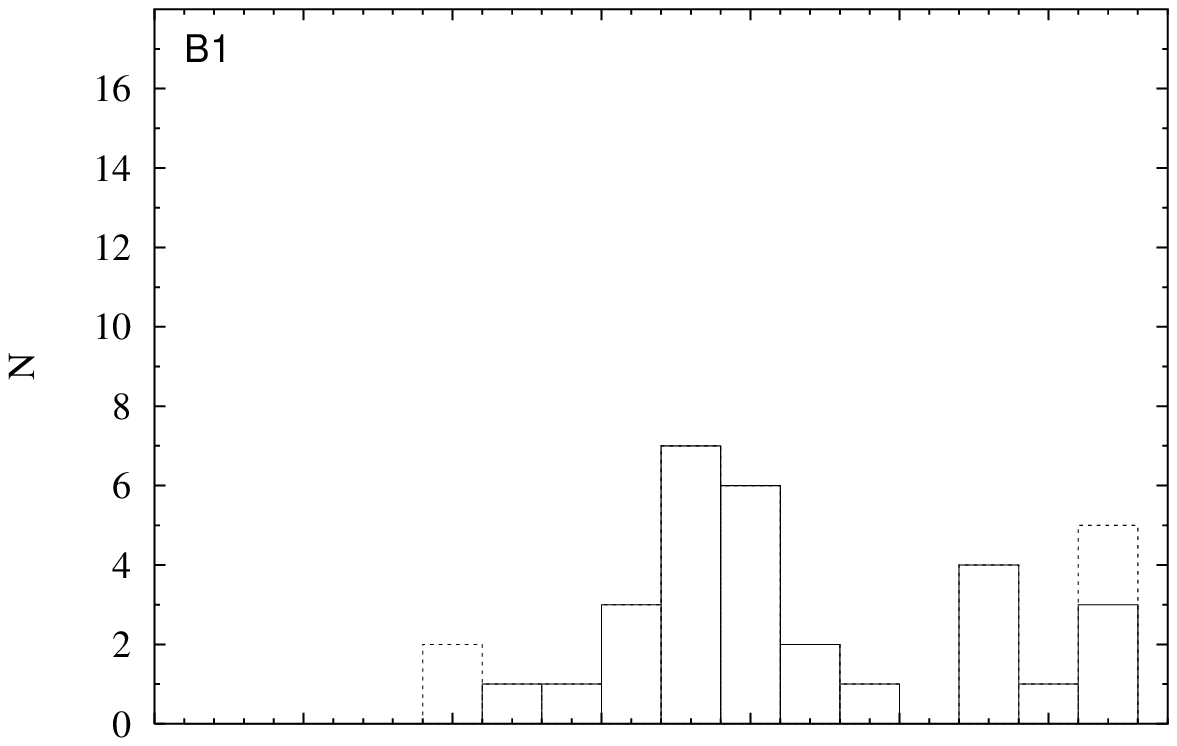, width=\linewidth}
\end{center}
\end{minipage}
\vfill
\vspace*{-9mm}
\begin{minipage}[b]{0.995\linewidth}
\begin{center}
\epsfig{figure=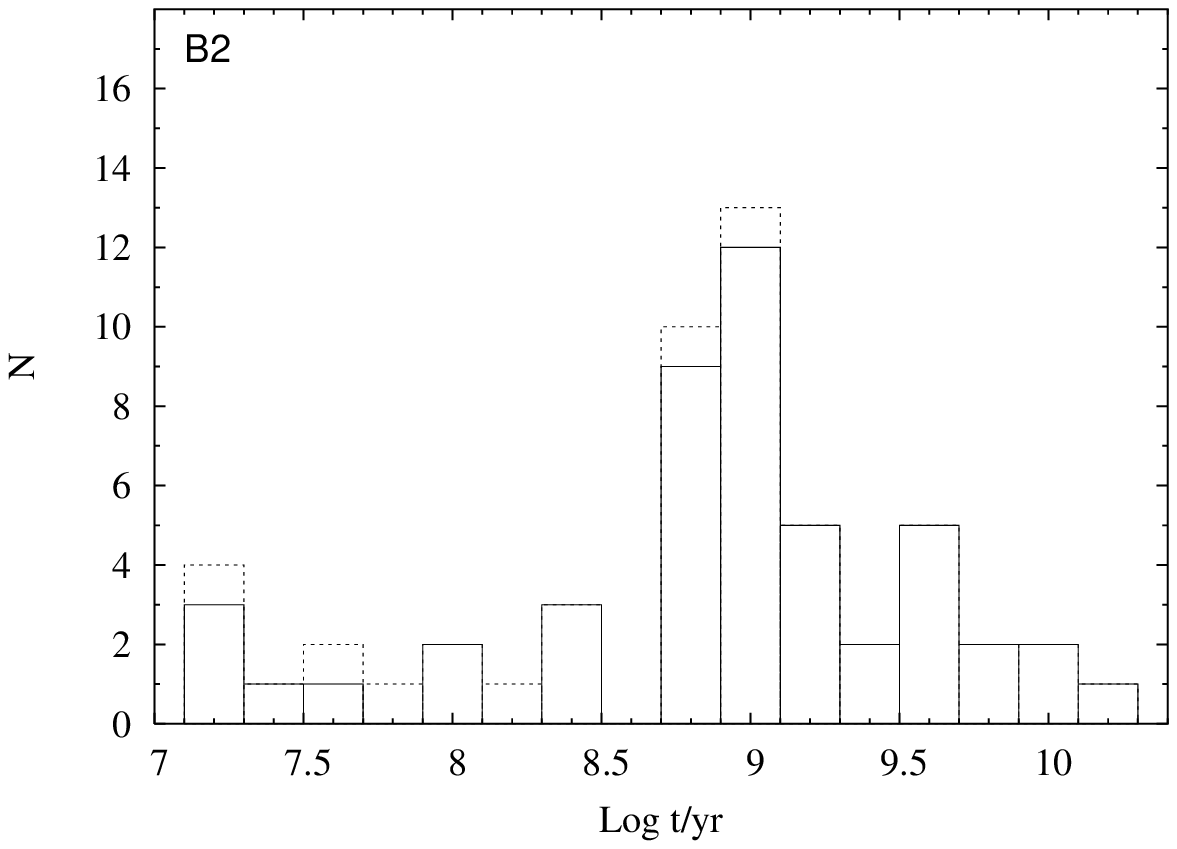, width=\linewidth}
\end{center}
\end{minipage}
\label{fig:ClFR}  
\caption{Age distribution of the star clusters in the entire M82~B region
(top panel), in the eastern half B1 (middle panel) and the western
half B2 (bottom panel).  Cluster age $t$ is expressed in years and 
dashed boxes represent clusters whose $B$, $V$ or $I$ is either 
an upper or lower limit}  
\end{figure}

\begin{figure}
\begin{minipage}[b]{0.995\linewidth}
\begin{center}
\epsfig{figure=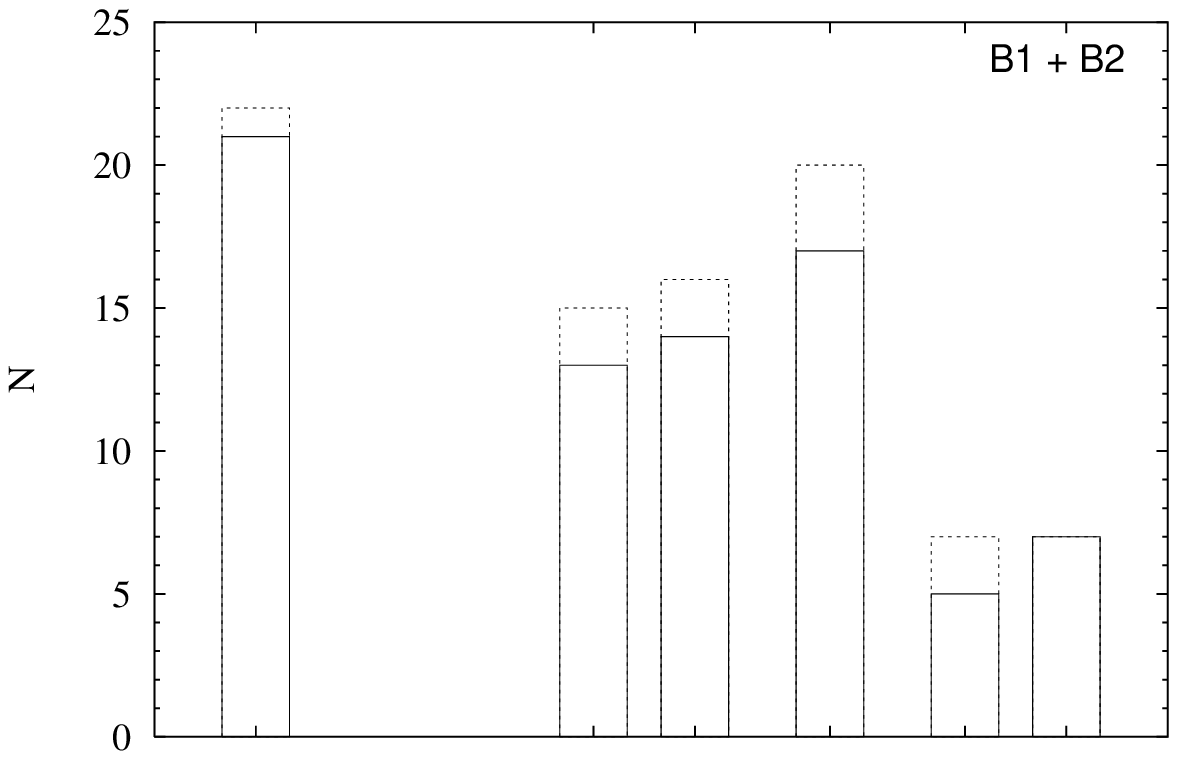, width=\linewidth}
\end{center}
\end{minipage}
\vfill
\vspace*{-9mm}
\begin{minipage}[b]{0.995\linewidth}
\begin{center}
\epsfig{figure=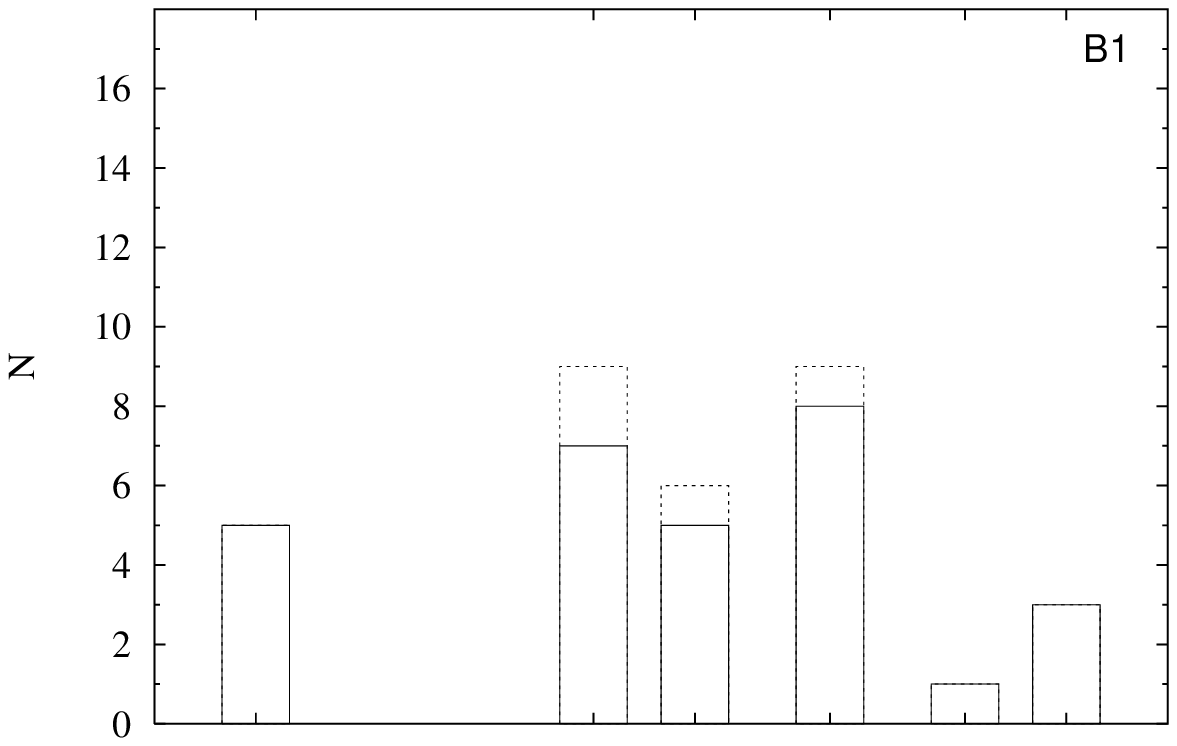, width=\linewidth}
\end{center}
\end{minipage}
\vfill
\vspace*{-9mm}
\begin{minipage}[b]{0.995\linewidth}
\begin{center}
\epsfig{figure=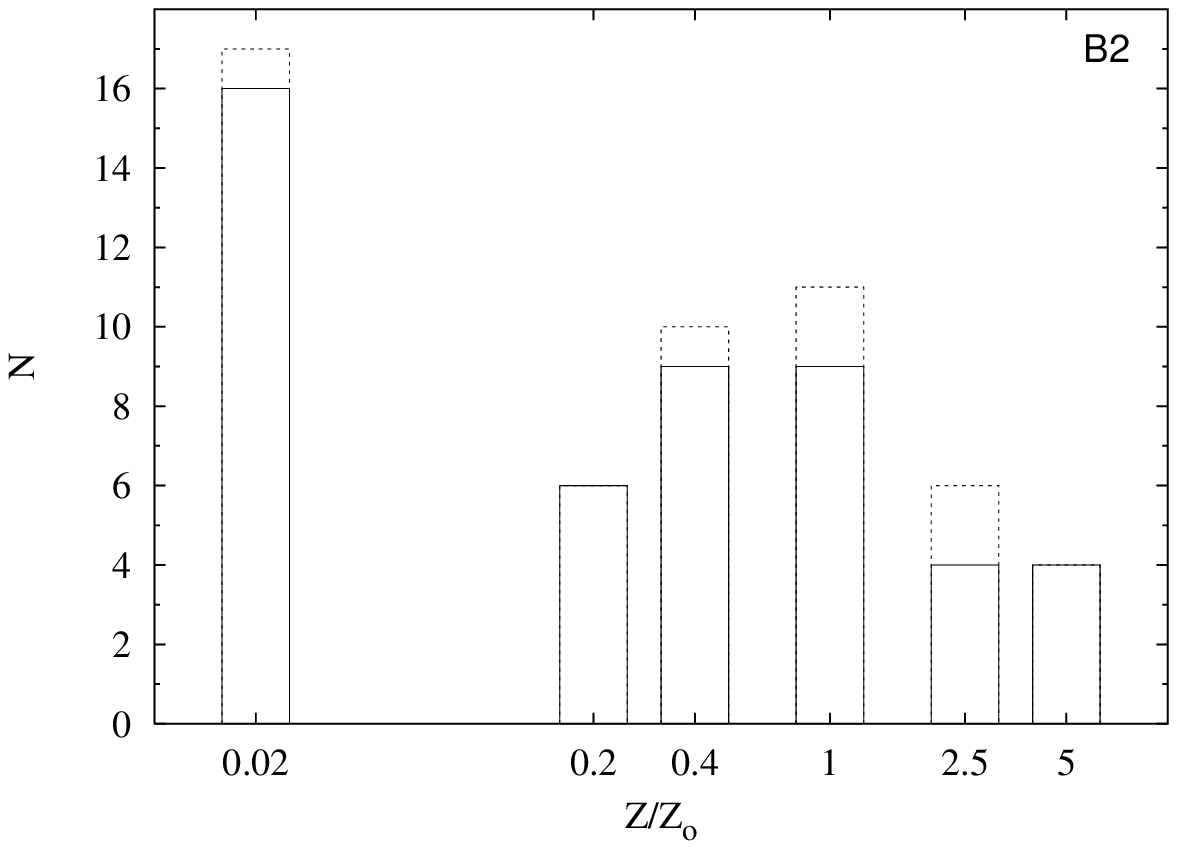, width=\linewidth}
\end{center}
\end{minipage}
\label{fig:ClFR}  
\caption{Metallicity distribution of the star clusters in the entire 
M82~B region (top panel), in the eastern half B1 (middle panel) and 
the western half B2 (bottom panel).  Dashed boxes represent clusters whose 
$B$, $V$ or $I$ is either an upper or lower limit}  
\end{figure}

\begin{figure}
\begin{minipage}[b]{0.98\linewidth}
\epsfig{figure=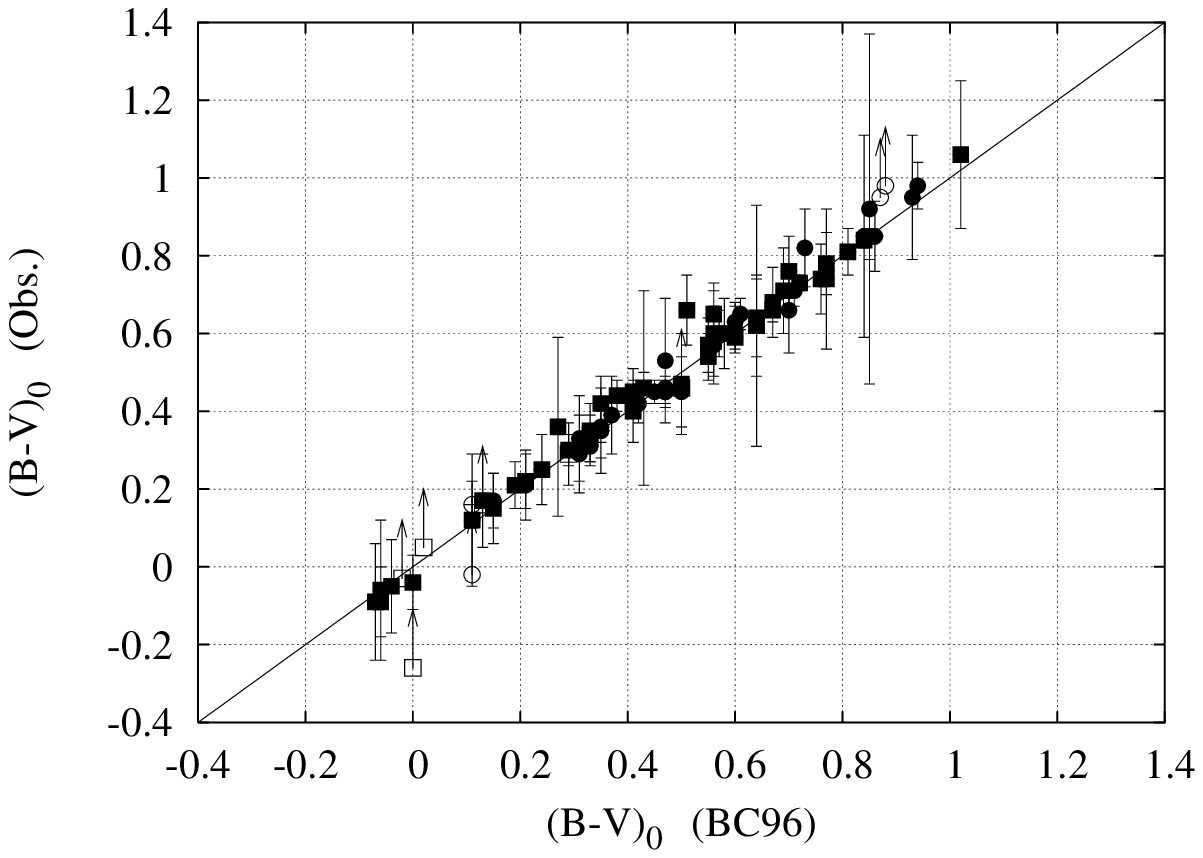, width=\linewidth} 
\end{minipage}
\vfill
\begin{minipage}[b]{0.98\linewidth}
\centering\epsfig{figure=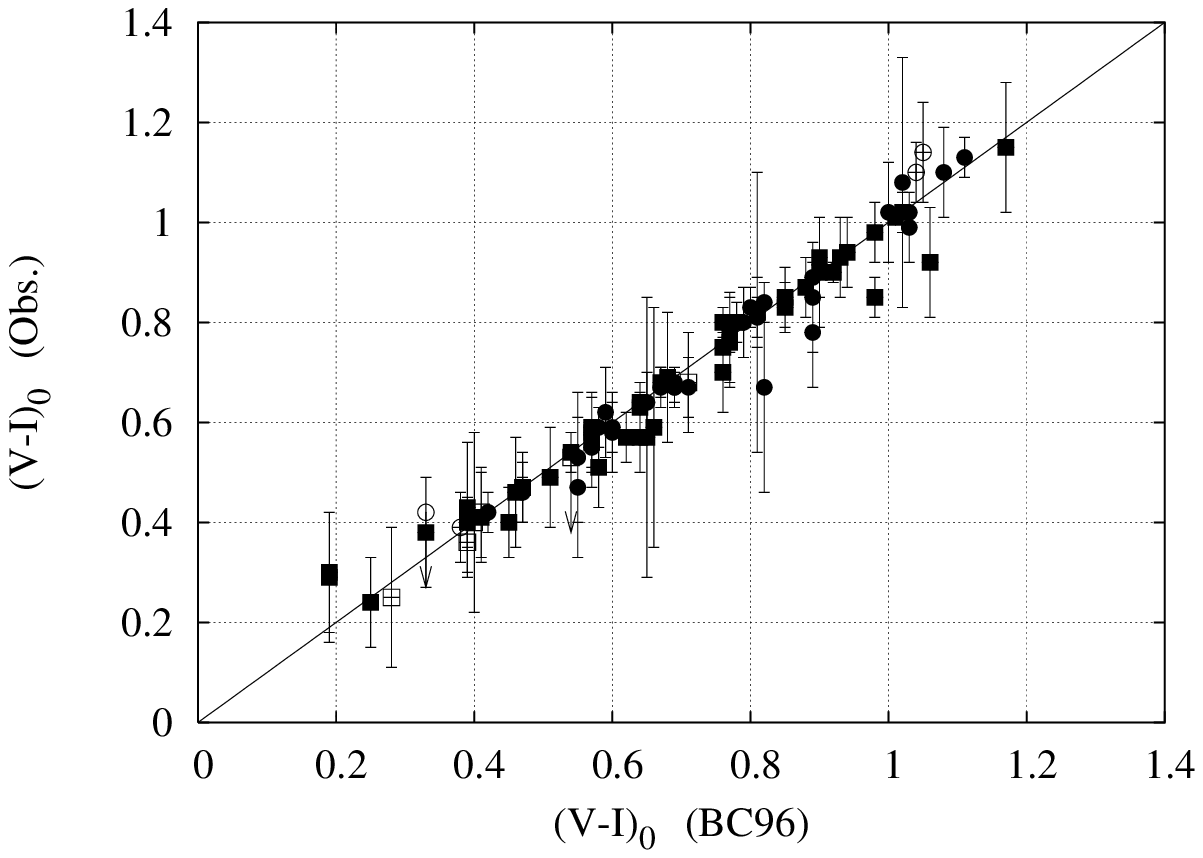, width=\linewidth}
\end{minipage}
\vfill
\begin{minipage}[b]{0.98\linewidth}
\centering\epsfig{figure=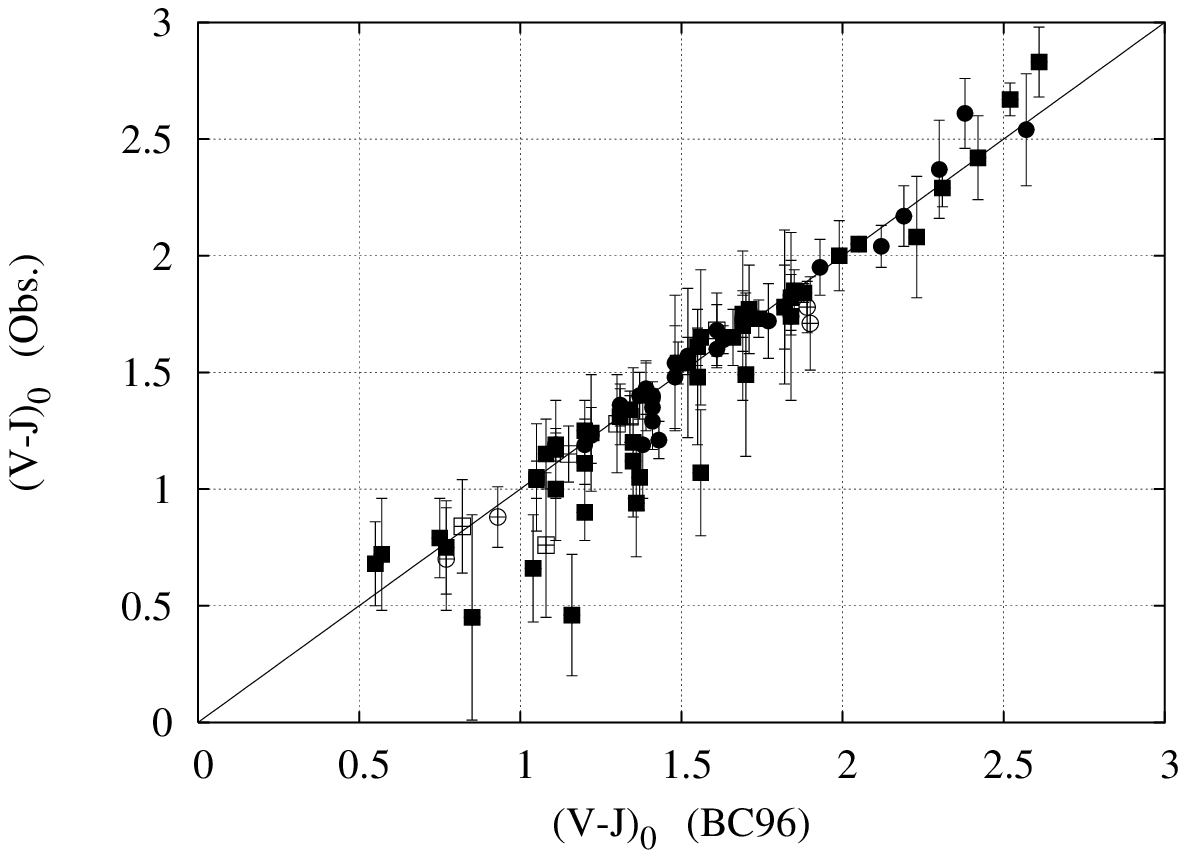, width=\linewidth}
\end{minipage}
\caption{Comparison between the intrinsic (i.e. observed and dereddened) 
colour indices and those of the BC96 model at the metallicities and 
ages determined for the B1/B2 clusters (circles/squares).  The 1$\sigma$ 
error bars are also indicated.  When either $B$, $V$ or $I$ is an upper or 
lower limit, the cluster is represented by an open symbol}
\label{fig:compareCI} 
\end{figure}

Figure 1 displays the reddening vector corresponding to an extinction of one
magnitude in the $VIJ$ diagram, i.e. the diagram used to 
disentangle the age and the metallicity.  Since this vector is roughly parallel 
to the iso-metallicity tracks, the extinction uncertainties (Eq.~(\ref{sig_Av})) will 
significantly propagate as age uncertainties.  An overestimate of the extinction 
leads to an underestimate of the cluster age, and vice versa.  In addition, the
age dependent mass-to-light ratio causes the age uncertainties to propagate 
as mass uncertainties. 

In order to perform a consistent estimate of the age uncertainties, we propagate the 
photometric uncertainties in each filter $B$, $V$, $I$ and $J$.  
The age $t$ of a cluster is a function of its $BVIJ$ photometry, that is, 
$({\rm log}\,t)~=~{\rm log}\,t(B,V,I,J)$.  Assuming that $B$,$V$,$I$ and $J$ 
are independent measurements, $\sigma _{{\rm log}\,t}$, the $1\sigma$
error in ${\rm log}\,t$ we aim to estimate, is given by:
\begin{multline}
\label{sig_logt1}
\sigma _{{\rm log}\,t}^2 
= \sigma _B ^2 \left( \frac{\partial ({\rm log}\,t)}{\partial B}\right)_{V,I,J}^2
+ \sigma _V ^2 \left( \frac{\partial ({\rm log}\,t)}{\partial V}\right)_{B,I,J}^2 \\
+ \sigma _I ^2 \left( \frac{\partial ({\rm log}\,t)}{\partial I}\right)_{B,V,J}^2
+ \sigma _J ^2 \left( \frac{\partial ({\rm log}\,t)}{\partial J}\right)_{B,V,I}^2\;.
\end{multline}
In this equation, $\sigma _B$ is the $B$ magnitude uncertainty, etc, 
$(\partial ({\rm log}\,t)/\partial B)_{V,I,J}$ is the partial derivative 
of the age logarithm with respect to $B$ while keeping $V$,$I$,$J$ constant, etc.  
Equation (\ref{sig_logt1}) can be rewritten as:
\begin{multline}
\sigma _{{\rm log}\,t}^2 = [\delta _B({\rm log}\,t)]^2 
+ [\delta _V({\rm log}\,t)]^2 \\
+ [\delta _I({\rm log}\,t)]^2 + [\delta _J({\rm log}\,t)]^2
\label{sig_Logt2}
\end{multline} 
where $\delta _B({\rm log}\,t)$ is the variation in ${\rm log}\,t$ caused by a 
1$\sigma$ change in the $B$ magnitude while keeping the $V$, $I$, $J$ ones 
to their best values, etc.  The different terms of the right-hand side of 
Eq.~(\ref{sig_Logt2}) can therefore be derived by computing the cluster age 
for each of the following combinations of observed magnitudes: 
$({\rm log}\,t)(B, V, I, J)$, 
$({\rm log}\,t)(B \pm \sigma _B, V, I, J)$,
$({\rm log}\,t)(B, V \pm \sigma _V, I, J)$, 
$({\rm log}\,t)(B, V, I \pm \sigma _I, J)$, 
$({\rm log}\,t)(B, V, I, J \pm \sigma _J)$, 
$B$, $V$, $I$ and $J$ being the best magnitude values.  
Figure 5 shows how such 1$\sigma$ individual changes in the observed magnitudes
of a cluster affect its location in the $VIJ$ diagram.  The left and right panels 
illustrate the cases of B1-43 and B2-32, respectively.  The crossed-circle indicates
the location of the cluster corresponding to its best magnitude values.  
The diagrams illustrate that changing these values by the following amounts:
\begin{itemize}
\item [a) ] $+ \sigma _B$ or $- \sigma _V$ or $+ \sigma _I$ or $+ \sigma _J$ 
(corresponding to the stars in the panels of Fig.~5) leads
to an increase of the age estimate.  Equation (\ref{Av_coef}) indeed indicates that
increasing $B$ or $I$, or decreasing $V$ leads to a decrease of the extinction
and, therefore, to a shift towards the right part of the $VIJ$ diagram, i.e. towards
older ages.  Increasing the $J$ magnitude does not affect the extinction but leads 
nevertheless to an age increase through a downwards shift in the $VIJ$ diagram;  
\item [b) ] a similar reasoning shows that changing the magnitudes by 
$- \sigma _B$ or $+ \sigma _V$ or $- \sigma _I$ or $- \sigma _J$ (the diamonds 
in the panels of Fig.~5) with respect of their best values lowers the cluster
age estimate. 
\end{itemize}
Summing quadratically the age changes derived from the series (a) will
therefore provide the age ``upper'' error bar, while summing quadratically the 
age changes derived from the series (b) will provide the ``lower'' age error bar, 
that is:
\begin{multline}
(+ \sigma _{{\rm log}\,t})^2 = 
[({\rm log}\,t)(B, V, I, J) - ({\rm log}\,t)(B + \sigma _B, V, I, J)]^2 \\
+ [({\rm log}\,t)(B, V, I, J) - ({\rm log}\,t)(B , V - \sigma _V, I, J)]^2 \\
+ [({\rm log}\,t)(B, V, I, J) - ({\rm log}\,t)(B , V, I + \sigma _I , J)]^2 \\
+ [({\rm log}\,t)(B, V, I, J) - ({\rm log}\,t)(B , V, I, J + \sigma _J )]^2
\end{multline}
and  
\begin{multline}
(- \sigma _{{\rm log}\,t})^2 = 
[({\rm log}\,t)(B, V, I, J) - ({\rm log}\,t)(B - \sigma _B, V, I, J)]^2 \\
+ [({\rm log}\,t)(B, V, I, J) - ({\rm log}\,t)(B , V + \sigma _V, I, J)]^2 \\
+ [({\rm log}\,t)(B, V, I, J) - ({\rm log}\,t)(B , V, I - \sigma _I , J)]^2 \\
+ [({\rm log}\,t)(B, V, I, J) - ({\rm log}\,t)(B , V, I, J -\sigma _J)]^2
\end{multline}
for the ``upper'' and ``lower'' error bars in ${\rm log}\,t$, respectively. \\

\begin{figure*}
\begin{minipage}[b]{0.49\linewidth}
\centering\epsfig{figure=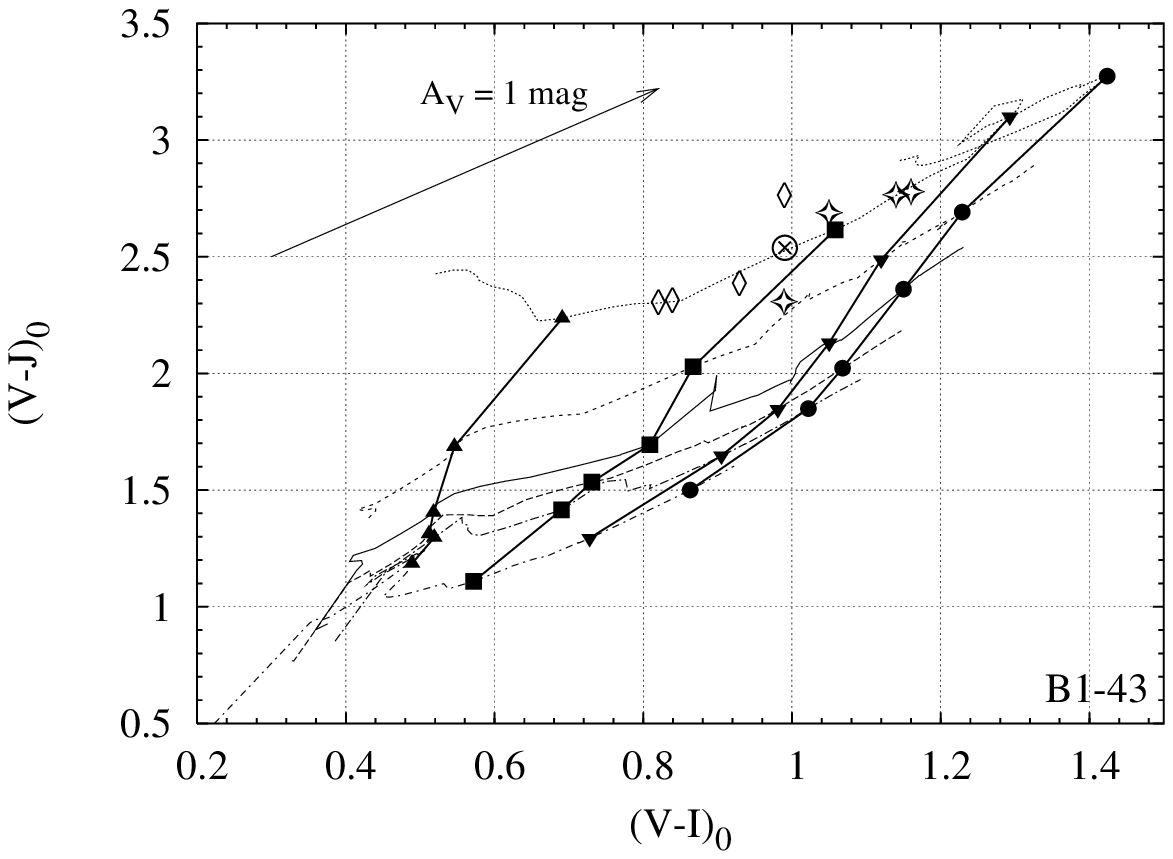, width=\linewidth} 
\end{minipage}
\hfill
\begin{minipage}[b]{0.49\linewidth}
\centering\epsfig{figure=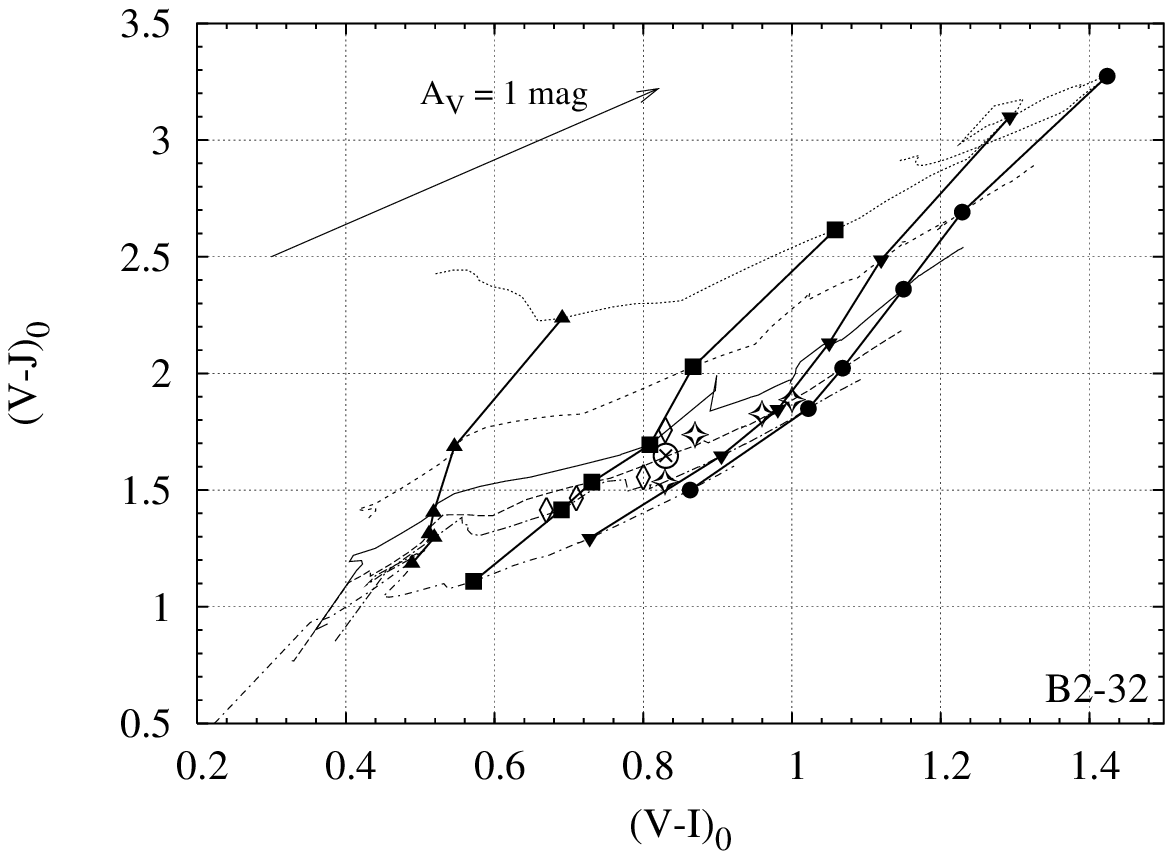, width=\linewidth}
\end{minipage}
\caption{Impact of the individual propagation of the $B$, $V$, $I$ and $J$ magnitude 
uncertainties upon the cluster location in the $VIJ$ diagram.  The crossed circle
stands for the location corresponding to the best magnitude values, the four-branch stars
correspond to the best values shifted by  $+ \sigma_B$ or$- \sigma_V$ or $+ \sigma_I$ 
or $+ \sigma_J$, and the diamonds are the cluster locations if the best values
have been shifted by $- \sigma_B$ or$+ \sigma_V$ or $- \sigma_I$ or $- \sigma_J$.
Left panel: B1-43; right panel: B2-32}
\label{fig:prop_er} 
\end{figure*}

As already mentioned, the age uncertainties propagate as mass uncertainties
through the age dependent mass-to-light ratio.  The errors affecting the cluster 
mass and metallicity are derived in a way similar to those affecting the age: we individually 
propagate the magnitude uncertainties and we then compute the lower/upper error 
bars by summing quadratically the shifts leading to mass and metallicity estimates 
smaller/greater than the best ones.  The 1$\sigma$ mass error we derive thus takes
into account the uncertainties affecting both the intrinsic luminosity of the 
cluster and its mass-to-light ratio.  The luminosity uncertainties originate from 
\begin{enumerate}
\item the visual magnitude uncertainty ($\sigma _V$), 
\item the extinction uncertainty, which depends on the optical magnitude
uncertainties, i.e. $\sigma _B$, $\sigma _V$ and $\sigma _I$ (see Eq.~(\ref{sig_Av})).
\end{enumerate}
The mass-to-light ratio uncertainties are driven by the age and 
metallicity uncertainties, which both depend on the $BVIJ$ photometry.  
The metallicity uncertainties however mostly reflect the $J$ magnitude 1$\sigma$ error.  
More than two thirds of our clusters are characterized by:
\begin{multline}
Z_{\rm max}=Z(B , V, I, J - \sigma _J) ~~ {\rm and} \\
Z_{\rm min}=Z(B , V, I, J + \sigma _J)\;.
\end{multline}  
This result is not unexpected: the extinction vector being roughly parallel 
to the aging trajectories, the influence of the extinction uncertainties (driven
by the optical uncertainties) upon the metallicity is strongly reduced. \\

The age, mass and metallicity estimates (minimal, best and maximal values) 
are listed in Tables 1 and 2 for B1 and B2, respectively.  Our photometric mass 
estimates show an accuracy of well within an order of magnitude.  Tables 1 and 2
show that the mass estimates are more robust than the age estimates 
(see also Fig.~8).  This pattern comes from the combination of two effects.
Firstly, the mass-to-light ratio grows with time at a slower rate than time 
itself (i.e. the evolution of ${\rm log}\,(m/L_v)$ vs ${\rm log}\,t$ exhibit 
a linear behaviour with a slope of less than unity, see BC96).  In addition, 
the impact upon the mass estimate of mass-to-light ratio variations is partially 
counterbalanced by extinction variations.
In order to illustrate this, let us consider a cluster for which the extinction
has been overestimated.  The adjustment of the extinction to a more appropriate 
(i.e. lower) value leads to a smaller cluster intrinsic brightness which,
alone, would lower the photometric mass estimate.  At the same time however,
the extinction decrease causes a rise of the cluster age estimate 
(see the extinction vector in Fig.~5) and, therefore, an increase of the 
cluster mass-to-light ratio.  These effects, a luminosity decrease and a 
mass-to-light ratio increase, act upon the mass estimate in opposite ways.
For instance, considering a roughly solar metallicity model, a decrease of
0.6\,magnitude in $A_V$ causes an age increase of ${\rm log}\,t \simeq 0.5$ 
at the time of the burst  (see Fig.~5).  Accordingly, the mass-to-light 
ratio grows by a factor of $\simeq$ 2.7.  The new mass estimate is thus larger 
than the previous one by a factor 2.7/2.5$^{0.6} \simeq 1.6$.  In summary, 
while an age increase leads to a growth of the mass estimate through the
age dependent mass-to-light ratio, this mass increase is reduced by a
decrease in extinction, which acts in the opposite way.  
It is worth to mention that this extinction-induced reduction of the 
mass variation is not so important for intermediate and old populations.  
In fact, due to the tightening of the isochrones at older age
(see Fig.~5), a given extinction variation corresponds to a
greater age range at older age than at the burst epoch and, therefore,
to a larger mass-to-light ratio variation.  The  relative reduction of the mass 
uncertainty thanks to the extinction variation is thus much weaker at old age
than at the burst age. 
  
As a result of the variations with time of the mass-to-light ratio,
the lower and upper mass limits are strongly coupled to the lower 
and upper age limits, respectively.
      
\begin{table*}
\begin{center}
\caption[]{Derived properties of the cluster sample in M82~B1, from left to right
the age, the mass, the metallicity and the extinction}
\begin{tabular}{rrrrcrclcccccccc} \hline \hline
\vspace{4pt}
ID~~~~ & \multicolumn{3}{c}{log\,$t$/yr} & & \multicolumn{3}{c}{log\,$m$/M$_{\odot}$} & &
\multicolumn{3}{c}{Z/Z$_{\odot}$} & & \multicolumn{3}{c}{A$_V$} \\ 
%ID~~~~ &  & ~log $t$/yr & & &  log\,$m$/M$_{\odot}$  &  & & Z/Z$_{\odot}$ & & & A$_V$ &  \\
% & min & best & Max &  min & best & Max & min & best & Max & min & best & Max \\
\hline 
1...~~~  & 8.92 & 9.80 & 10.18 &  & 5.03 & 5.58 & 5.88 & & 0.20 & 0.40 & 1.00 & & 0.92 & 1.30 & 1.38 \\
4...~~~  & 7.81 & 8.41 &  8.89 &  & 4.98 & 5.24 & 5.43 & & 1.00 & 1.00 & 2.50 & & 2.00 & 2.48 & 2.69 \\
5...~~~  & 9.98 & 10.18 & 10.18 & & 5.86 & 6.01 & 6.01 & & 0.20 & 0.20 & 0.40 & & 0.42 & 0.73 & 1.04 \\
6...~~~  & 7.91 & 8.21 & 8.83 & & 5.26 & 5.34 & 5.57 & & 0.02 & 1.00 & 1.00 & & 1.47 & 1.86 & 1.97 \\
7...~~~  & 9.03 & 9.20 & 9.37 & & 5.57 & 5.68 & 5.78 & & 0.20 & 0.40 & 0.40 & & 0.22 & 0.51 & 0.80 \\
8...~~~  & 8.69 & 8.86 & 8.98 & & 5.24 & 5.29 & 5.36 & & 0.02 & 0.02 & 0.20 & &0.12 & 0.35 & 0.58 \\
9...~~~  & 8.33 & 8.71 & 8.82 & & 4.20 & 4.30 & 4.43 & & 5.00 & 5.00 & 5.00 & & 0.00 & 0.00 & 0.73 \\
11...~~~ & 8.89 & 9.01 & 9.17 & & 5.35 & 5.40 & 5.49 & & 0.02 & 0.20 & 0.20 & & 0.06 & 0.34 & 0.62 \\
12...~~~ & 8.72 & 9.02 & 9.31 & & 4.78 & 4.84 & 5.03 & & 0.20 & 1.00 & 2.50 & & 0.00 & 0.23 & 0.97 \\
14...~~~ & 8.37 & 8.78 & 9.05 & & 5.04 & 5.25 & 5.38 & & 0.40 & 1.00 & 2.50 & & 0.89 & 1.39 & 1.89 \\
16...~~~ & 8.60 & 10.18 & 10.18 & &  4.87 & 5.76 & 5.76 & & 0.02 & 0.20 & 0.40 & & 1.71 & 1.71 & 2.88 \\
17...~~~ & 8.37 & 8.81 & 9.16 & & 4.50 & 4.63 & 4.73 & & 5.00 & 5.00 & 5.00 & & 0.01 & 0.67 & 1.33 \\
18...~~~ & 8.85 & 8.98 & 9.05 & & 5.51 & 5.56 & 5.60 & & 0.20 & 0.20 & 0.40 & & 0.40 & 0.59 & 0.78 \\
20...~~~ & 9.50 & 9.83 & 10.18 & & 5.85 & 6.00 & 6.19 & & 0.02 & 0.02 & 0.02 & & 0.09 & 0.32 & 0.55 \\
22...~~~ & 8.37 & 8.71 & 8.83 & & 5.15 & 5.30 & 5.36 & & 1.00 & 1.00 & 2.50 & & 0.70 & 0.98 & 1.26 \\
24...~~~ & 8.68 & 8.90 & 9.06 & & 5.13 & 5.20 & 5.32 & & 0.20 & 0.20 & 0.40 & & 0.20 & 0.57 & 0.94 \\
25...~~~ & 9.30 & 9.80 & 10.18 & & 5.28 & 5.54 & 5.82 & & 0.02 & 0.02 & 0.02 & & 0.48 & 0.80 & 1.12 \\
26...~~~ & 8.22 & 8.66 & 8.77 & & 4.14 & 4.33 & 4.38 & & 0.20 & 1.00 & 2.50 & & 0.00 & 0.00 & 0.23 \\
28...~~~ & 8.96 & 9.00 & 9.08 & & 5.95 & 5.98 & 6.01 & & 0.20 & 0.20 & 0.20 & & 0.46 & 0.55 & 0.64 \\
29...~~~ & 7.82 & 8.75 & 8.98 & & 4.20 & 4.72 & 4.75 & & 0.02 & 0.40 & 1.00 & & 0.45 & 1.07 & 1.69 \\
30...~~~ & 8.98 & 9.20 & 9.36 & & 5.03 & 5.13 & 5.20 & & 0.02 & 0.02 & 0.20 & & 0.08 & 0.29 & 0.50 \\
32...~~~ & 8.94 & 9.74 & 10.18 & & 4.68 & 5.10 & 5.42 & & 0.02 & 0.02 & 0.02 & & 0.00 & 0.34 & 0.85 \\
33...~~~ & 7.10 & 10.18 & 10.18 & & 3.51 & 5.01 & 5.01 & & 0.02 & 0.20 & 0.40 & & 0.00 & 0.20 & 2.45 \\
36...~~~ & 8.60 & 10.00 & 10.18 & & 4.86 & 5.91 & 6.10 & & 0.20 & 1.00 & 1.00 & & 0.85 & 1.01 & 1.83 \\
37...~~~ & 9.13 & 9.38 & 10.07 & & 5.70 & 5.82 & 6.26 & & 0.20 & 0.20 & 0.40 & & 0.63 & 0.86 & 1.09 \\
38...~~~ & 8.10 & 8.68 & 8.92 & & 4.37 & 4.67 & 4.71 & & 0.40 & 2.50 & 2.50 & & 0.13 & 0.75 & 1.37 \\
40...~~~ & 7.73 & 8.50 & 8.95 & & 3.70 & 3.90 & 4.13 & & 0.20 & 1.00 & 2.50 & & 0.00 & 0.54 & 1.29 \\
41...~~~ & 8.22 & 8.75 & 8.90 & & 4.79 & 4.97 & 5.01 & & 0.20 & 0.40 & 1.00 & & 0.63 & 1.03 & 1.43 \\
43...~~~ & 8.62 & 8.90 & 9.28 & & 4.31 & 4.44 & 4.59 & & 2.50 & 5.00 & 5.00 & & 0.00 & 0.48 & 1.03 \\
\hline \hline
\end{tabular}
\end{center}
\label{tab:B1}
%\vspace{2cm}
\end{table*}

\begin{table*}
\begin{center}
\caption[]{Derived properties of the cluster sample in M82~B2, from left to right
the age, the mass, the metallicity and the extinction}
\begin{tabular}{rrrrcrclcccccccc} \hline \hline
\vspace{4pt}
ID~~~~ & \multicolumn{3}{c}{log\,$t$/yr} & & \multicolumn{3}{c}{log\,$m$/M$_{\odot}$} & &
\multicolumn{3}{c}{Z/Z$_{\odot}$} & & \multicolumn{3}{c}{A$_V$} \\ 
%ID~~~~ &  & ~log $t$/yr & & &  log\,$m$/M$_{\odot}$  &  & & Z/Z$_{\odot}$ & & & A$_V$ &  \\
% & min & best & Max &  min & best & Max & min & best & Max & min & best & Max \\
\hline 
 1...~~~  & 8.68 & 10.09 & 10.15 & & 4.35 & 5.25 & 5.31 & & 1.00 & 1.00 & 2.50 & & 0.00 & 0.00 & 0.64 \\
 4...~~~  & 8.79 &  8.91 &  8.98 & & 5.41 & 5.45 & 5.46 & & 5.00 & 5.00 & 5.00 & & 0.49 & 0.73 & 0.96 \\
 5...~~~  & 7.49 &  8.76 &  9.20 & & 4.89 & 5.38 & 5.47 & & 0.02 & 0.02 & 0.40 & & 0.75 & 1.34 & 1.92 \\
 8...~~~  & 8.73 &  9.01 &  9.62 & & 5.55 & 5.59 & 6.00 & & 0.40 & 1.00 & 2.50 & & 2.72 & 3.35 & 3.98 \\
12...~~~  & 9.27 &  9.44 &  9.58 & & 6.60 & 6.71 & 6.82 & & 1.00 & 1.00 & 1.00 & & 0.83 & 0.94 & 1.06 \\
13...~~~  & 9.36 &  9.98 & 10.18 & & 5.38 & 5.77 & 5.88 & & 0.20 & 0.20 & 0.40 & & 0.54 & 0.85 & 1.16 \\
14...~~~  & 8.31 &  9.01 &  9.81 & & 4.78 & 5.05 & 5.56 & & 5.00 & 5.00 & 5.00 & & 1.00 & 1.83 & 2.66 \\
15...~~~  & 8.87 &  8.96 &  9.08 & & 5.88 & 5.98 & 6.00 & & 0.20 & 1.00 & 1.00 & & 2.88 & 3.20 & 3.52 \\
17...~~~  & 8.93 &  9.60 & 10.18 & & 5.27 & 5.62 & 6.04 & & 0.20 & 0.20 & 0.40 & & 0.62 & 1.18 & 1.74 \\   
18...~~~  & 7.10 &  7.16 &  8.18 & & 4.64 & 4.64 & 5.29 & & 0.02 & 0.20 & 0.40 & & 3.89 & 4.40 & 4.70 \\
21...~~~  & 7.09 &  7.31 &  7.72 & & 5.06 & 5.16 & 5.42 & & 2.50 & 2.50 & 2.50 & & 4.11 & 4.18 & 4.98 \\
23...~~~  & 7.21 &  9.01 & 10.18 & & 4.25 & 4.75 & 5.77 & & 0.02 & 1.00 & 2.50 & & 0.00 & 1.70 & 3.24 \\
25...~~~  & 7.77 &  8.41 &  9.12 & & 4.72 & 4.95 & 5.25 & & 0.20 & 0.40 & 1.00 & & 0.61 & 1.26 & 1.70 \\
26...~~~  & 8.86 &  9.01 &  9.13 & & 5.65 & 5.72 & 5.77 & & 0.02 & 0.02 & 0.20 & & 0.48 & 0.58 & 0.68 \\ 
28...~~~  & 7.01 &  7.18 &  8.00 & & 3.67 & 4.03 & 4.62 & & 0.02 & 0.20 & 0.40 & & 1.93 & 2.30 & 3.14 \\
29...~~~  & 7.18 &  8.01 &  9.74 & & 3.53 & 3.81 & 4.54 & & 0.02 & 0.20 & 0.40 & & 0.29 & 1.06 & 1.95 \\
30...~~~  & 8.79 &  9.00 &  9.29 & & 5.15 & 5.21 & 5.22 & & 0.02 & 0.02 & 0.20 & & 1.14 & 1.56 & 1.98 \\
32...~~~  & 8.87 &  9.16 &  9.80 & & 5.09 & 5.19 & 5.62 & & 0.02 & 0.40 & 1.00 & & 1.20 & 1.66 & 2.13 \\
33...~~~  & 7.38 &  8.46 &  9.16 & & 3.75 & 4.30 & 4.51 & & 2.50 & 5.00 & 5.00 & & 0.00 & 1.13 & 2.12  \\
34...~~~  & 7.20 &  8.31 &  8.83 & & 3.57 & 4.09 & 4.22 & & 0.02 & 1.00 & 2.50 & & 0.76 & 1.29 & 1.52  \\
36...~~~  & 9.27 &  9.30 &  9.68 & & 5.35 & 5.37 & 5.60 & & 0.20 & 0.20 & 0.40 & & 0.00 & 0.00 & 0.02  \\
37...~~~  & 7.53 &  7.60 &  7.75 & & 5.19 & 5.25 & 5.34 & & 0.40 & 2.50 & 2.50 & & 2.43 & 2.43 & 3.58 \\
38...~~~  & 8.90 &  9.16 &  9.30 & & 4.52 & 4.76 & 4.86 & & 0.02 & 0.02 & 0.20 & & 0.00 & 0.00 & 0.00 \\
39...~~~  & 7.84 &  8.81 &  9.02 & & 4.02 & 4.42 & 4.56 & & 0.20 & 1.00 & 2.50 & & 0.00 & 0.00 & 0.80 \\
40...~~~  & 8.69 &  8.81 &  8.90 & & 5.23 & 5.27 & 5.31 & & 0.02 & 0.02 & 0.02 & & 0.42 & 0.54 & 0.66 \\
41...~~~  & 8.77 &  8.86 &  8.98 & & 5.89 & 5.91 & 5.96 & & 2.50 & 2.50 & 2.50 & & 0.54 & 0.74 & 0.92 \\
43...~~~  & 9.06 &  9.06 &  9.35 & & 4.54 & 4.54 & 4.84 & & 0.02 & 0.02 & 0.02 & & 0.00 & 0.00 & 0.00 \\
45...~~~  & 8.77 &  9.65 & 10.18 & & 4.94 & 5.44 & 5.81 & & 0.02 & 0.02 & 0.02 & & 1.25 & 1.54 & 2.33 \\
47...~~~  & 9.50 & 10.18 & 10.18 & & 4.64 & 5.00 & 5.00 & & 0.02 & 0.02 & 0.02 & & 0.00 & 0.00 & 0.45 \\
48...~~~  & 8.33 &  8.81 &  9.05 & & 4.23 & 4.46 & 4.52 & & 0.02 & 0.02 & 1.00 & & 0.56 & 1.04 & 1.18 \\
49...~~~  & 9.09 &  9.51 & 10.18 & & 5.49 & 5.71 & 6.28 & & 0.20 & 0.40 & 1.00 & & 0.23 & 0.63 & 1.02 \\
50...~~~  & 7.17 &  7.20 &  8.81 & & 4.06 & 4.09 & 4.82 & & 0.02 & 0.40 & 0.40 & & 1.39 & 2.12 & 2.50 \\
52...~~~  & 9.50 &  9.70 &  9.85 & & 4.89 & 5.07 & 5.16 & & 0.02 & 0.02 & 0.02 & & 0.00 & 0.00 & 0.07 \\
54...~~~  & 8.15 &  8.96 &  9.08 & & 4.21 & 4.55 & 4.57 & & 0.02 & 0.02 & 0.40 & & 0.00 & 0.01 & 0.54 \\
55...~~~  & 8.60 &  8.71 &  9.07 & & 3.93 & 3.98 & 4.07 & & 0.02 & 1.00 & 2.50 & & 0.00 & 0.00 & 0.00 \\    
56...~~~  & 8.71 &  9.11 &  9.16 & & 4.84 & 4.91 & 4.97 & & 1.00 & 1.00 & 2.50 & & 0.00 & 0.11 & 0.83 \\
57...~~~  & 8.46 &  8.76 &  9.05 & & 5.15 & 5.28 & 5.38 & & 2.50 & 5.00 & 5.00 & & 0.63 & 1.13 & 1.63 \\
59...~~~  & 6.70 &  8.06 &  9.72 & & 3.99 & 4.65 & 5.34 & & 0.02 & 0.40 & 1.00 & & 1.93 & 2.88 & 3.81 \\
60...~~~  & 7.96 &  9.01 &  9.70 & & 4.14 & 4.60 & 4.91 & & 0.02 & 0.02 & 0.20 & & 0.23 & 0.77 & 1.33 \\
62...~~~  & 8.98 &  9.65 &  9.78 & & 4.08 & 4.47 & 4.57 & & 0.02 & 0.02 & 0.20 & & 0.00 & 0.00 & 0.38 \\
63...~~~  & 8.70 &  8.96 &  9.10 & & 4.66 & 4.83 & 4.83 & & 0.02 & 0.02 & 0.40 & & 0.49 & 0.75 & 0.99 \\
64...~~~  & 9.40 &  9.68 & 10.18 & & 4.40 & 4.58 & 4.98 & & 0.02 & 0.20 & 0.40 & & 0.00 & 0.00 & 0.34 \\
65...~~~  & 9.14 &  9.72 & 10.17 & & 5.16 & 5.51 & 5.72 & & 0.02 & 0.02 & 0.02 & & 0.52 & 0.84 & 1.16 \\
66...~~~  & 8.20 &  8.76 &  8.99 & & 3.80 & 3.95 & 3.99 & & 0.40 & 2.50 & 2.50 & & 0.00 & 0.16 & 1.06 \\
67...~~~  & 8.76 &  8.91 &  9.01 & & 5.21 & 5.23 & 5.28 & & 0.20 & 0.40 & 1.00 & & 1.21 & 1.45 & 1.67 \\
68...~~~  & 8.73 &  9.23 &  9.59 & & 4.52 & 4.69 & 4.96 & & 0.02 & 0.40 & 1.00 & & 0.00 & 0.39 & 0.90 \\
69...~~~  & 8.72 &  9.06 &  9.81 & & 4.01 & 4.16 & 4.60 & & 0.02 & 0.40 & 2.50 & & 0.00 & 0.00 & 0.00 \\
70...~~~  & 8.99 &  9.16 &  9.33 & & 4.95 & 5.12 & 5.19 & & 0.02 & 0.02 & 0.02 & & 0.05 & 0.17 & 0.29 \\ 
\hline \hline
\end{tabular}
\end{center}
\label{tab:B2}
%\vspace{2cm}
\end{table*}

%--------------------------------------------
\subsection{The cluster formation history}
%--------------------------------------------
Figure 2 shows the age distributions of the clusters
in the entire M82~B region (top panel), in regions B1 (middle panel) 
and B2 (bottom panel).  
Clusters have been forming continuously, their age ranging from more
than 10\,Gyr down to about 10\,Myr.  In addition, 
the B1 and B2 regions exhibit a period of enhanced cluster 
formation, from about 1.5\,Gyr ago until 500\,Myr ago.  We therefore 
confirm the main result of de Grijs et al.~(2001), although our peak
of cluster formation is slightly older, $\sim$ 1\,Gyr
(see also de Grijs et al. 2002, for more details).
The M82~B burst was followed by a sharp decline in the 
cluster formation rate.  Less than one fifth of the clusters are younger
than 500\,Myr.  Starbursts are likely to be strongly self-limited
by supernova-driven outflows which remove the remaining cool gas 
from the immediate starburst region.  
It is interesting to note that the B2 region, which
is closer to the active starburst, has formed a few clusters 
more recently than B1.    \\

It has long been suspected that the tidal interactions between the M81 group
members (i.e. M81, M82 and NGC~3077) are responsible for the starburst 
nature of M82 and for the HI bridges connecting these galaxies.
Brouillet et al.~(1991) used an N-body model, representing M81, M82 and 
NGC\,3077, to simulate the deformations 
of the neutral hydrogen distribution due to the tidal interactions: 
their results reproduce the observed HI bridges and tails rather well.
According to their simulation, the last perigalactic passage of M82
and M81 took place some 500\,Myr ago.  The onset of the burst
about 1.5\,Gyr ago is thought, therefore, to coincide with the beginning of
the tidal interactions between M82 and its prominent neighbour M81. 

%-----------------------------------------------------------
\subsection{The chemical evolution}
\begin{figure}
\begin{minipage}[b]{0.55\linewidth}
\hspace*{-2mm}
\epsfig{figure=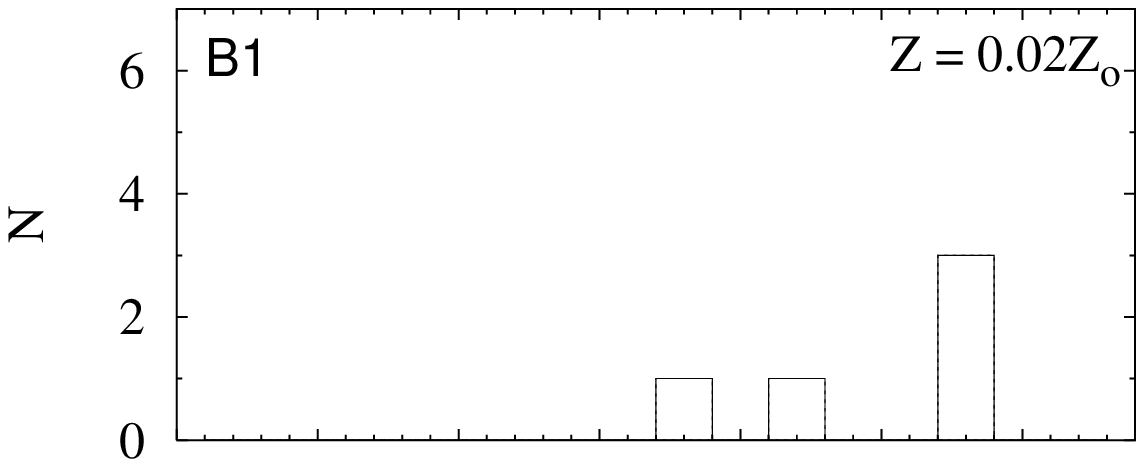, width=\linewidth}
\end{minipage}
\hfill \hspace*{-15mm} 
\begin{minipage}[b]{0.55\linewidth}
\hspace*{-2mm}
\epsfig{figure=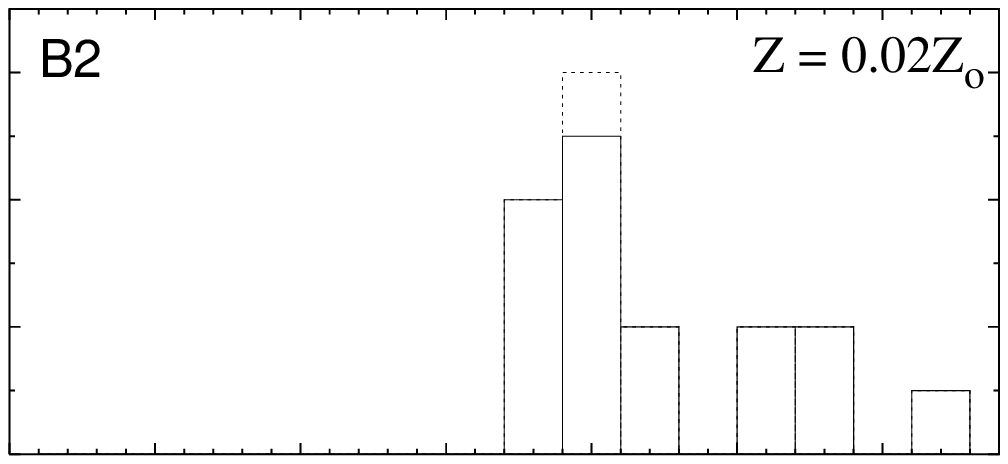, width=0.97\linewidth}
\end{minipage}
\vfill
\vspace*{-15mm}
\begin{minipage}[b]{0.55\linewidth}
\hspace*{-2mm}
\epsfig{figure=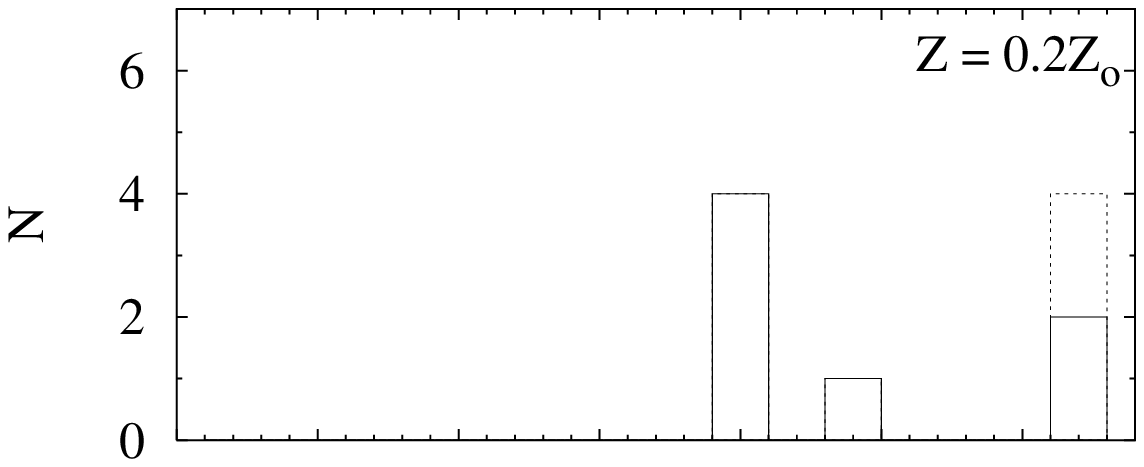, width=\linewidth}
\end{minipage}
\hfill \hspace*{-15mm}
\begin{minipage}[b]{0.55\linewidth}
\hspace*{-2mm}
\epsfig{figure=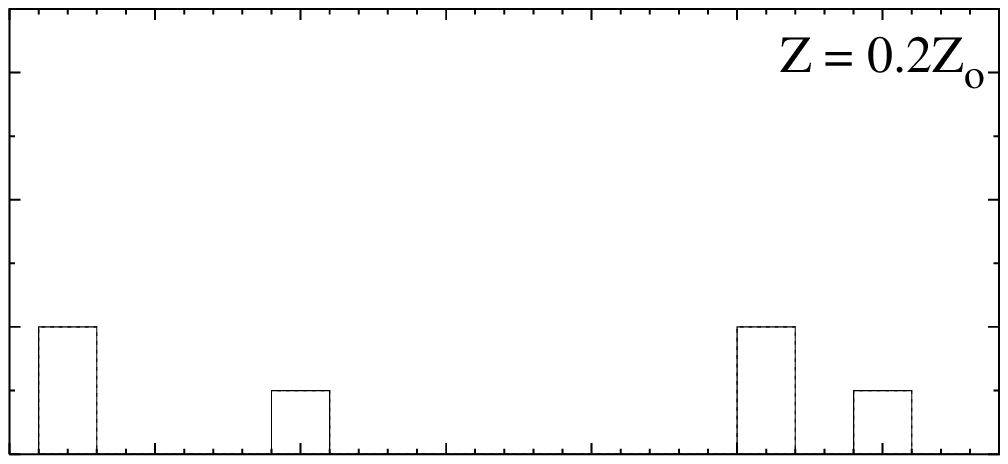, width=0.98\linewidth}
\end{minipage}
\vfill
\vspace*{-15.3mm}
\begin{minipage}[b]{0.55\linewidth}
\hspace*{-2mm}
\epsfig{figure=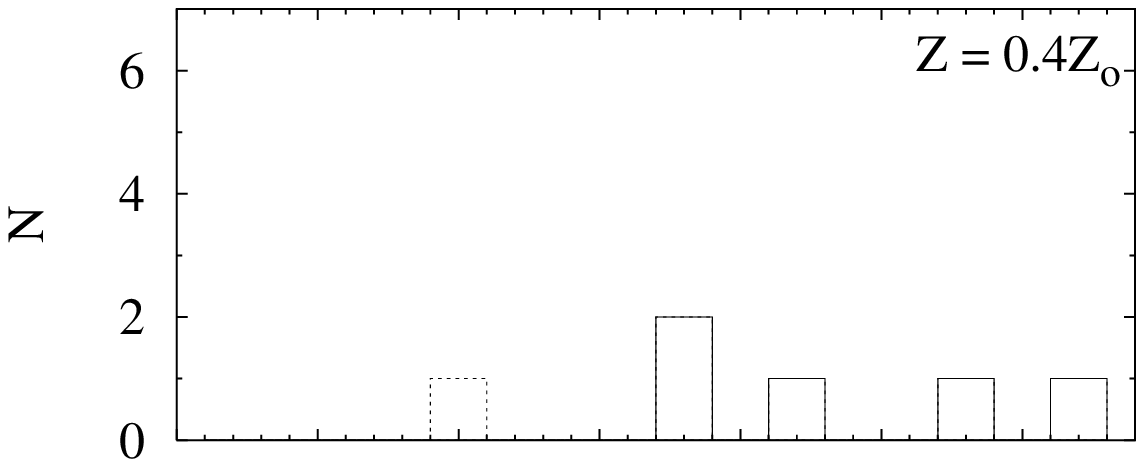, width=\linewidth}
\end{minipage}
\hfill \hspace*{-15mm}
\begin{minipage}[b]{0.55\linewidth}
\hspace*{-2mm}
\epsfig{figure=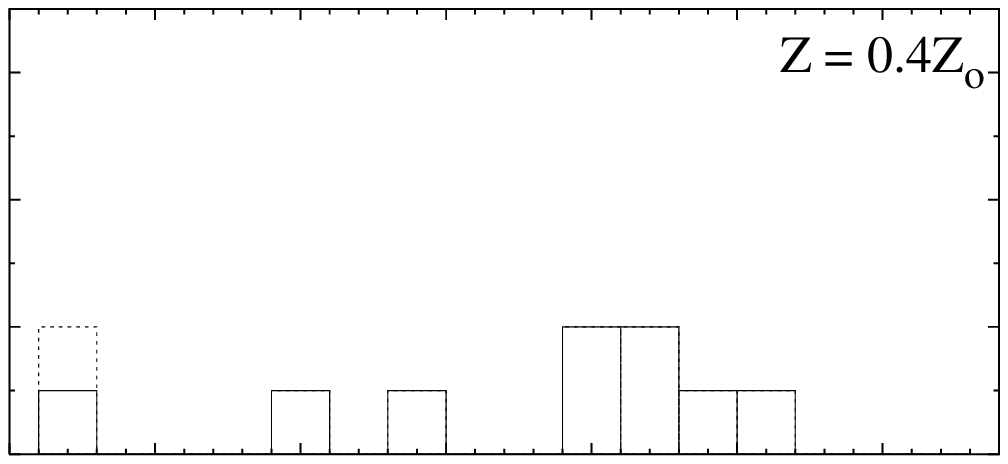, width=0.97\linewidth}
\end{minipage}
\vfill
\vspace*{-15mm}
\begin{minipage}[b]{0.55\linewidth}
\hspace*{-2mm}
\epsfig{figure=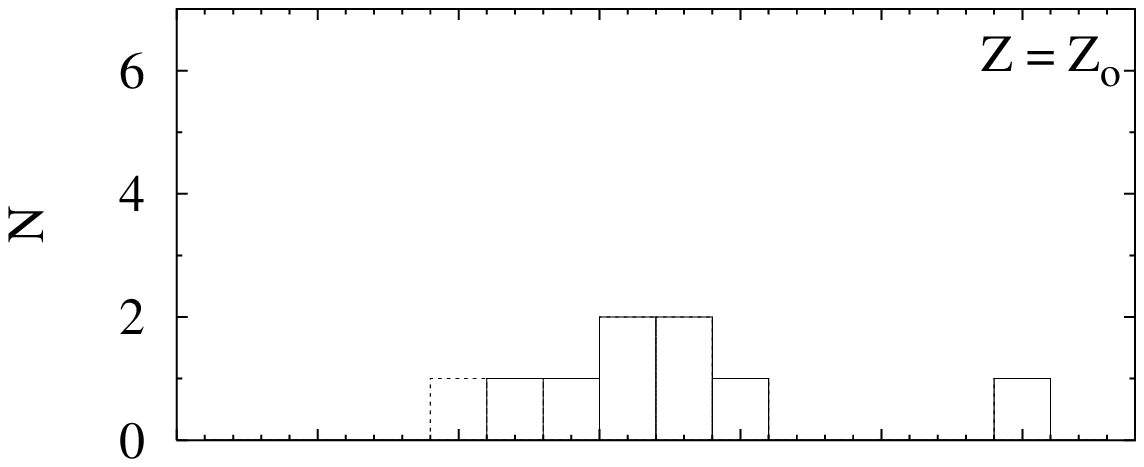, width=\linewidth}
\end{minipage}
\hfill \hspace*{-15mm}
\begin{minipage}[b]{0.55\linewidth}
\hspace*{-2mm}
\epsfig{figure=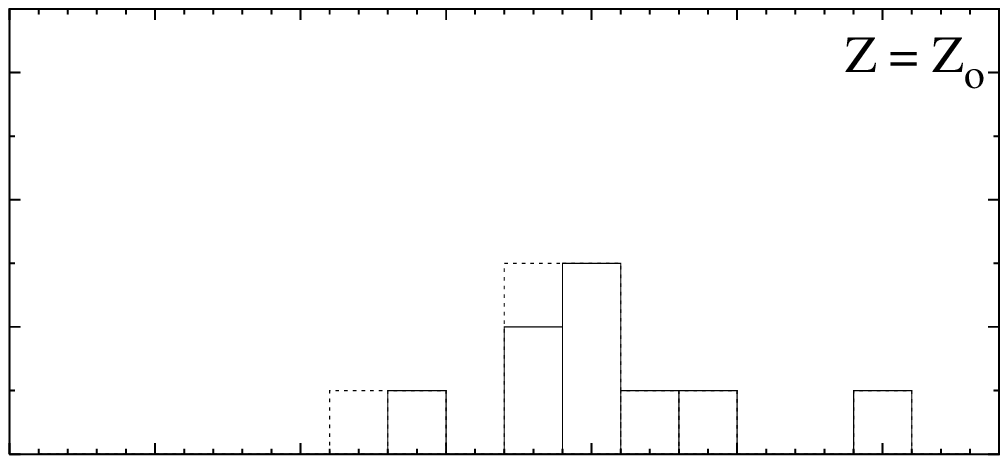, width=0.98\linewidth}
\end{minipage}
\vfill
\vspace*{-15mm}
\begin{minipage}[b]{0.55\linewidth}
\hspace*{-2mm}
\epsfig{figure=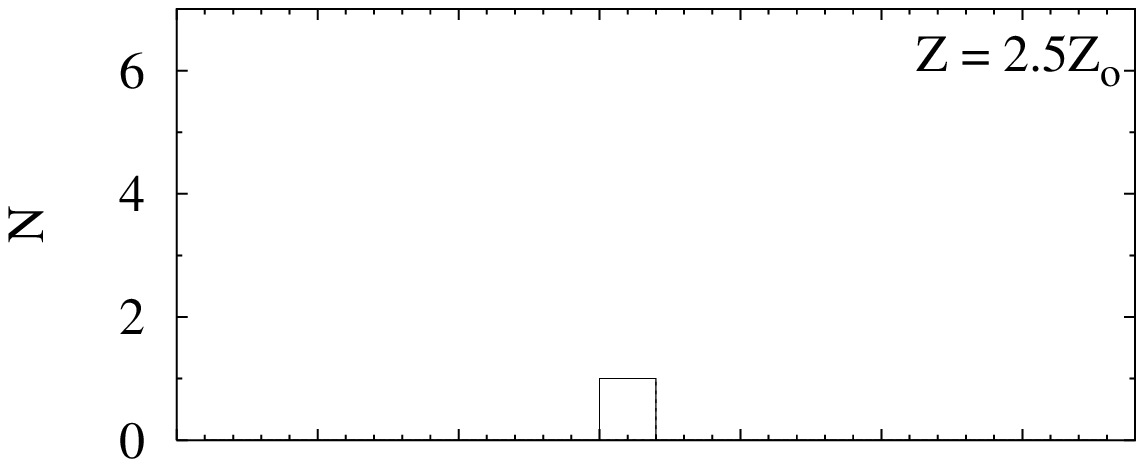, width=\linewidth}
\end{minipage}
\hfill \hspace*{-15mm}
\begin{minipage}[b]{0.55\linewidth}
\hspace*{-2mm}
\epsfig{figure=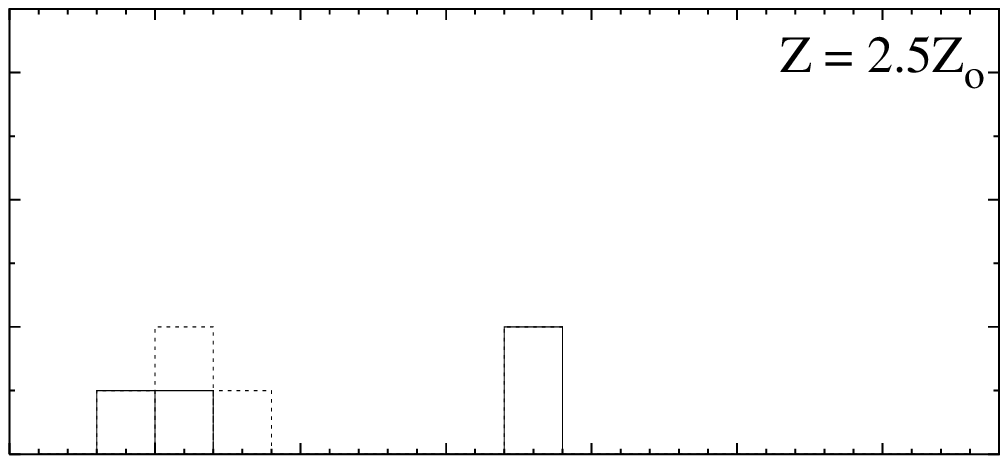, width=0.97\linewidth}
\end{minipage}
\vfill
\vspace*{-15mm}
\begin{minipage}[b]{0.55\linewidth}
\hspace*{-2mm}
\epsfig{figure=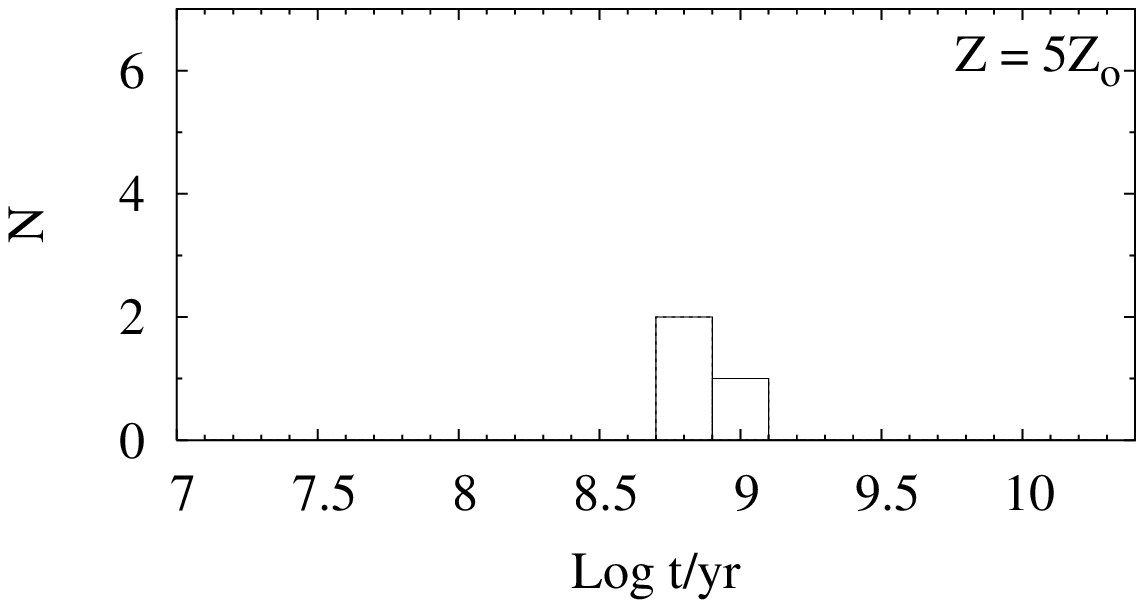, width=\linewidth}
\end{minipage}
\hfill \hspace*{-15mm}
\begin{minipage}[b]{0.55\linewidth}
\hspace*{-2mm}
\epsfig{figure=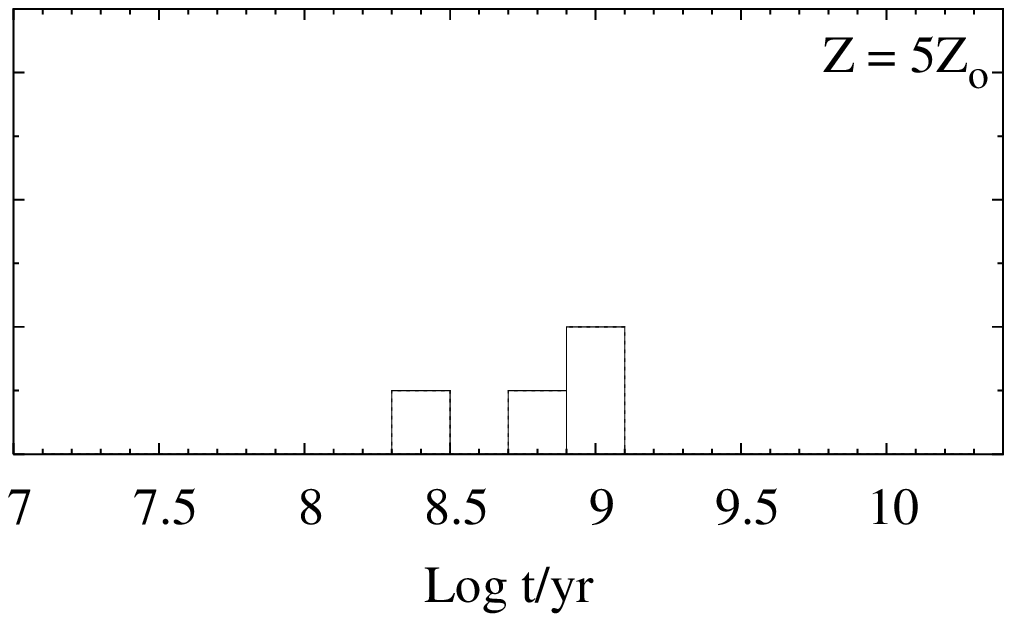, width=0.98\linewidth}
\end{minipage}
\caption{Age distributions of the M82~B clusters at
a given metallicity, from Z\,=\,0.02Z$_{\odot}$ (top) to Z\,=\,5Z$_{\odot}$
(bottom).  Left/right column: B1/B2 sample.  Dashed boxes represent  
clusters whose $B$, $V$ or $I$ is either an upper or lower limit}
\label{fig:histo_feh}
\end{figure}

%-----------------------------------------------------------
The M82~B cluster system is characterized by a wide range of metallicities
(see Fig.~3) as it hosts both very metal-poor clusters 
(Z$\simeq$0.02Z$_{\odot}$) and super-solar abundance ones
(Z$>$Z$_{\odot}$).  Such a pattern must be related
to the region's chemical evolution.  In order to illustrate this,
each panel of Fig.~\ref{fig:histo_feh} shows the cluster age distribution
at a given metallicity for regions B1 (left column) and B2 (right column). 
The metallicities shown are, from top to bottom, $Z = 0.02Z_{\odot}$,
$0.2Z_{\odot}$, $0.4Z_{\odot}$, $1Z_{\odot}$, $2.5Z_{\odot}$, $5Z_{\odot}$. \\

\begin{figure}
\begin{minipage}[b]{\linewidth}
\begin{flushright}
\epsfig{figure=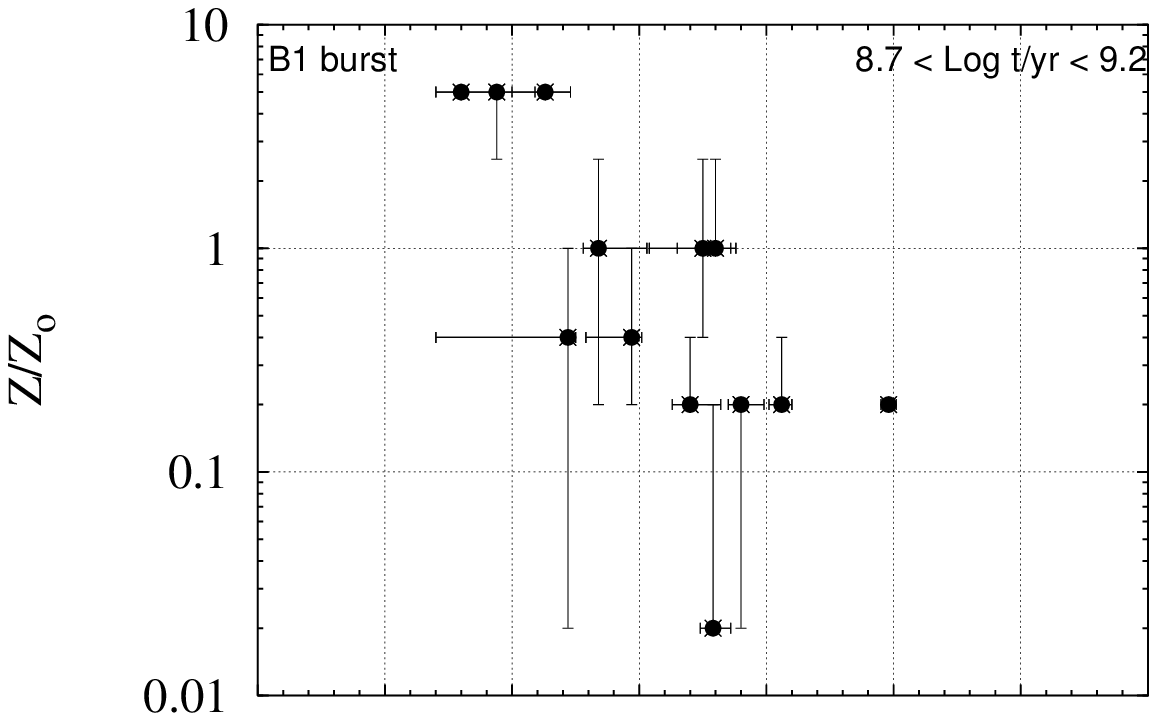, width=\linewidth}
\end{flushright}
\end{minipage}
\vfill \vspace*{-13mm}
\begin{minipage}[b]{\linewidth}
\begin{flushleft}
\epsfig{figure=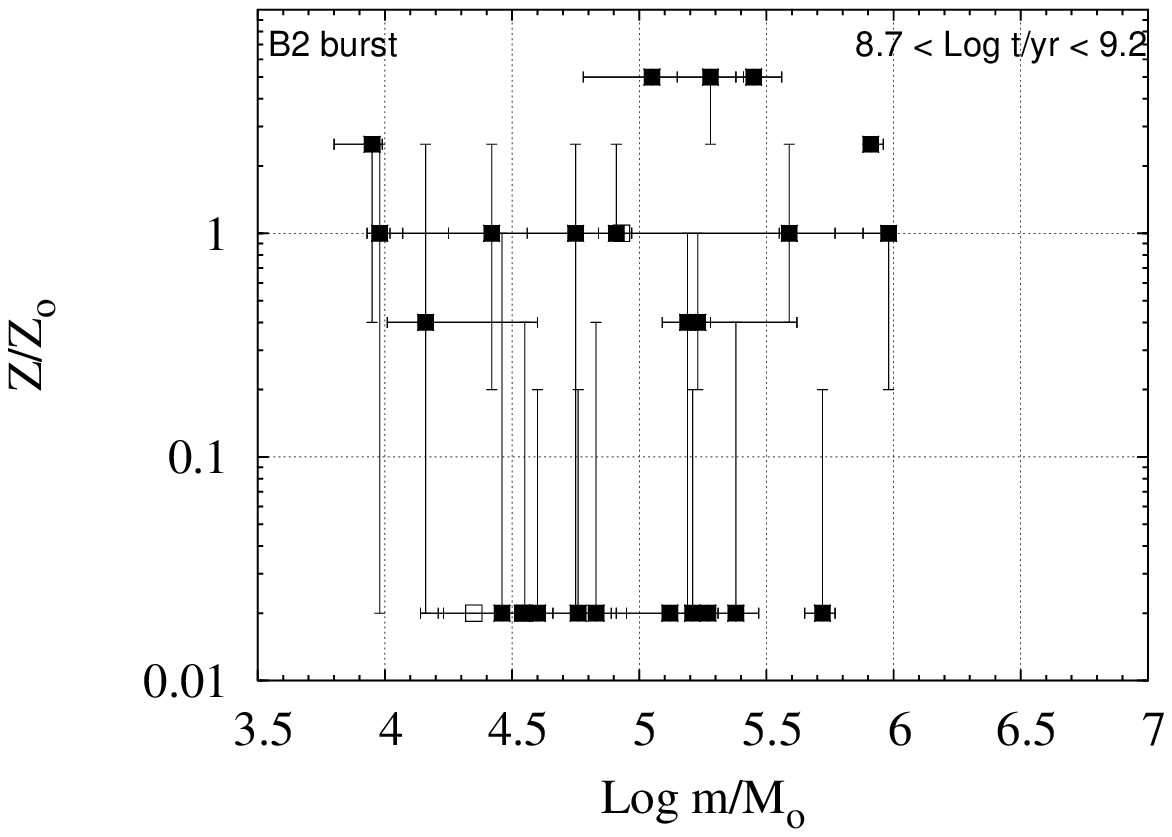, width=\linewidth}
\end{flushleft}
\end{minipage}
\caption{Mass-metallicity diagrams for the clusters formed during the
burst of cluster formation (0.5 $< t <$ 1.5\,Gyr).  The cluster mass $m$
is expressed in units of M$_{\odot}$.  Top/bottom panel: B1/B2.
Open symbols refer to clusters of which one of the observed magnitudes
$B$, $V$ or $I$ is ill-determined (upper or lower limit)}
\label{fig:ZLogm_Bb}
\end{figure} 

From the birth of its first clusters until about 1\,Gyr ago, M82~B formed 
clusters with roughly solar and subsolar metal abundances only.  In both 
subregions B1 and B2, the formation of the first clusters with significantly 
super-solar (i.e. $Z > Z_{\odot}$) metallicities coincides with the onset of 
the burst.  The stellar ejecta from the evolution of older stars have most 
probably contributed to the chemical enrichment. \\
While this chemical evolution up to super solar metallicities at the burst
epoch is a feature in common for B1 and B2, both subregions nevertheless
exhibit a striking difference regarding the most metal-poor clusters.
At the time of the burst, and despite the chemical enrichment noticed above,
B2 managed to form a significant subpopulation of very metal-poor clusters
(i.e. $Z \simeq 0.02Z_{\odot}$) with respect to B1 (top panels in Fig.~6).  
We stress that half of these B2 clusters have very well-constrained ages 
($({\rm log}\,t_{\rm max}-{\rm log}\,t_{\rm min}) \leq 0.5$) and metallicities
($Z_{\rm max} = 0.2$), so that we probably do not face a spurious effect.   

Either some interstellar gas in B2 managed to escape the chemical 
enrichment driven by the older stellar populations, or alternatively 
some ``fresh'' metal-poor gas was injected into the B2 region shortly 
before the burst.  Whereas the first scenario is hard to explain, the 
second is more appealing.  A large amount of circumgalactic cold material 
is orbiting M82, stripped from the gas-rich outskirts 
of the M81 group spirals by tidal interactions.  These interactions
involve quite a few galaxies, of which three are of a non-negligible size
(i.e. M81, M82 and NGC\,3077).  It is therefore not unreasonable 
to assume that tidal interactions and gas tidal stripping were at least
somewhat important even before M81 and M82 almost collided about 0.5\,Gyr 
ago.  If the bridges of circumgalactic gas were indeed already present 
1\,Gyr ago
(i.e. 500\,Myr prior to the perigalactic passage), gas infall from this
reservoir onto the B2 region may have induced the formation of this
subpopulation of metal-poor clusters.  Alternatively, some metal-poor 
gas associated with the M82 outskirts could have been driven onto B2 by 
tidal interactions. 

In comparing the properties of the clusters of B1 with those of B2, we 
implicitely assume that their respective current and birth locations, with 
respect to the galactic centre, 
coincide.  The validity of this assumption may be questioned as, at the 
distance from the centre of region B, one would expect M82's differential
rotation (Shen \& Lo 1995) to have caused the starburst area to disperse.  
However, the lack of bright clusters like those 
in M82~B outside this region does not meet this expectation and, on the 
contrary, indicates that the fossil starburst region has remained relatively 
well constrained.  The reason for this is likely found  in the complex structure 
of the disc.  It is well-known that the inner $\sim 1$ kpc of M82 
(i.e. the radial extent of M82~B) is 
dominated by a stellar bar (e.g., Wills et al.~2000) in solid-body rotation. 
From observations in other galaxies, it appears to be a common feature that 
central bars are often surrounded by a ring-like structure.  If this is also true 
for M82, it is reasonable to assume that stars in the ring are trapped, and
therefore cannot move very much in radius because of dynamical resonance
effects.  As a consequence, not only has the fossil starburst site managed
to stay together, both subregions B1 and B2 may also have conserved their own 
specificity for about 1\,Gyr (see also de Grijs 2001).  In the next section,
we show that the presence (B2), or the absence (B1), of a subpopulation of 
very metal-poor clusters at the burst epoch is not
the only difference between B1 and B2.   

%........................................................................
\subsubsection{Did the B1 clusters go through a self-enrichment phase?} 
%........................................................................
A high-pressure medium favours the formation of pressure-bound clouds, 
which may be the progenitors of future stellar clusters.  For instance, 
Fall \& Rees (1985) suggested that the formation of the halo GCs in our 
Galaxy occured in dense and cold clouds which are in pressure 
equilibrium with 
a hot and diffuse background.  This two-phase medium is thought to
have been formed by the collapse of the protoGalaxy.  
In the frame of their theory, the GC progenitor clouds are thermally
supported and made of primordial gas.   
To explain the metallicities of the Galactic halo GCs, 
Parmentier et al.~(1999) further developed this picture 
for GC formation and proposed that the halo GC gaseous progenitors 
went through a self-enrichment phase.  A self-enrichment scenario assumes 
the formation of a first generation of stars in the central regions
of each proto-globular cluster cloud.  When the massive stars explode
as Type II supernovae, they chemically enrich the surrounding cloud gas
and trigger the formation of a second stellar generation, more
metal-rich than the first one.  These second generation stars form the 
proto-globular 
cluster.  Supernovae having long been thought to disrupt the cloud of gas 
out of which they have formed, Parmentier et al.~(1999) studied the 
ability of pressure-truncated clouds to retain supernova ejecta and further
showed that such a process is able to explain Galactic halo GC 
metallicities.  Assuming that the number of supernovae 
is the maximum number predicted by their model (i.e. a few hundred), a 
self-enrichment episode in pressure-bound clouds shows up as a 
correlation between the mass and the achieved metallicity of the progenitor 
cloud in the sense that the least massive clouds are the most 
metal-rich.  Such a trend emerges because if the bound pressure is higher, 
the mass of the pressure-truncated cloud will be lower, and its ability 
to retain supernova ejecta will be greater.
Regarding the resulting system of GCs, a tight mass-metallicity 
correlation should not be expected, however.  The star formation 
efficiency (i.e. the ratio between the mass of the second stellar 
generation and the mass of gas) and the number of supernovae (which determines 
the amount of metals dispersed within the cloud) may vary
from cloud to cloud, imprinting therefore a scatter on the initial 
mass-metallicity correlation (see Parmentier \& Gilmore 2001 
for a more detailed discussion).  In spite of this scatter,    
Parmentier \& Gilmore (2001) showed that such a mass-metallicity
correlation is statistically present in the Milky Way Old Halo, 
i.e. the group of old and
coeval halo GCs (e.g. Rosenberg et al.~1999).  
At this stage, it is worth keeping in mind that the search for a 
mass-metallicity relation must be restricted to a coeval 
population of clusters.  The (surviving) correlation primarily reflects the
variations in the pressure by which the clouds are bound and has
nothing to do with the chemical enrichment with time. \\

Both the M82 interstellar medium and the hot protogalactic background
are high-pressure mediums.  Therefore, 
a tempting question is whether the almost coeval population of clusters formed 
during the burst (0.5 $< t <$ 1.5\,Gyr) have undergone a self-enrichment 
process in pressure-truncated progenitor clouds.  These M82~B clouds 
were most likely mainly supported by non-thermal effects (e.g. turbulence 
and magnetic fields) and resembled more closely giant molecular clouds
than the neutral hydrogen clouds described by the Fall \& Rees (1985)
theory.  As mentioned above, however, the key parameter leading to a
trend between the mass and the metallicity of stellar clusters is the 
pressure of the medium in which their progenitor clouds were embedded.
It is widely believed that the boundary of giant molecular clouds is
set by pressure balance with the surrounding, more diffuse, 
interstellar medium (e.g. McLaughlin \& Pudritz 1996).  Therefore,   
the hypothesis of a mass-metallicity relation among the clusters of the 
M82~B system makes sense.
In order to explore the possibility of a self-enrichment event
(superimposed on an unavoidable pre-enrichment phase owing to the young
age of the clusters), we now search for such a mass-metallicity correlation 
among the $\simeq$1\,Gyr old clusters in the B1 and B2 regions.
Again, B1 and B2 exhibit different patterns (see Fig.~\ref{fig:ZLogm_Bb},
B1/B2: top/bottom panel).
The B1 sample shows a mass-metallicity correlation in the sense expected by
the self-enrichment model, i.e. the most metal-rich clusters are, 
on average, least massive: the linear Pearson correlation coefficient is  
$r=-$0.70, corresponding to a correlation probability of $\wp$ = 99.5\%.
The scatter in metallicity at a given mass interval 
(e.g. $10^5 < m < 3\times 10^5 M_{\odot}$) remains significant but this is
not unexpected because of the scatter sources mentioned above.  The age of the
vast majority of the B1 burst clusters is well constrained 
($({\rm log}\,t_{\rm max}-{\rm log}\,t_{\rm min}) \leq 0.7$ for 11 of the 13 clusters)
and so is their mass (see the horizontal error bars in the top panel of Fig.~7).
In addition, the metallicity error bars are rather limited as well.
As a result, the error bars, both in mass and metallicity, do not 
significantly corrupt the correlation.    \\
In contrast to B1, the B2 burst clusters do not exhibit any 
mass-metallicity correlation.  
Removing the most metal-poor clusters (i.e. those assumed to have 
formed from circumgalactic gas) does not change the situation. 
This difference between the B1 and B2 mass-metallicity diagrams 
might be related to the formation
of some metal-poor clusters in B2 at the burst epoch.  If  
these low metal abundance clusters indeed formed out of the arrival of 
``fresh'' gas, a shock wave would have accompanied the gas injection,
perturbing the external pressure and preventing therefore any 
self-enrichment in pressure-bound clouds. \\
While the scenario presented above is tentative, the fact that B1
and B2 exhibit different behaviours from the points of view of the
chemical evolution (top panels of Fig.~\ref{fig:histo_feh}) and of the 
mass-metallicity relation (Fig.~\ref{fig:ZLogm_Bb}) is puzzling and suggests 
actual differences between the cluster formation histories in both regions
at the time of the burst.  \\

%-------------------------------------------------------------------
\subsection{Dynamical disruption of the SSCs}
%------------------------------------------------------------------
\begin{figure}
\begin{minipage}[b]{\linewidth}
\begin{flushright}
\epsfig{figure=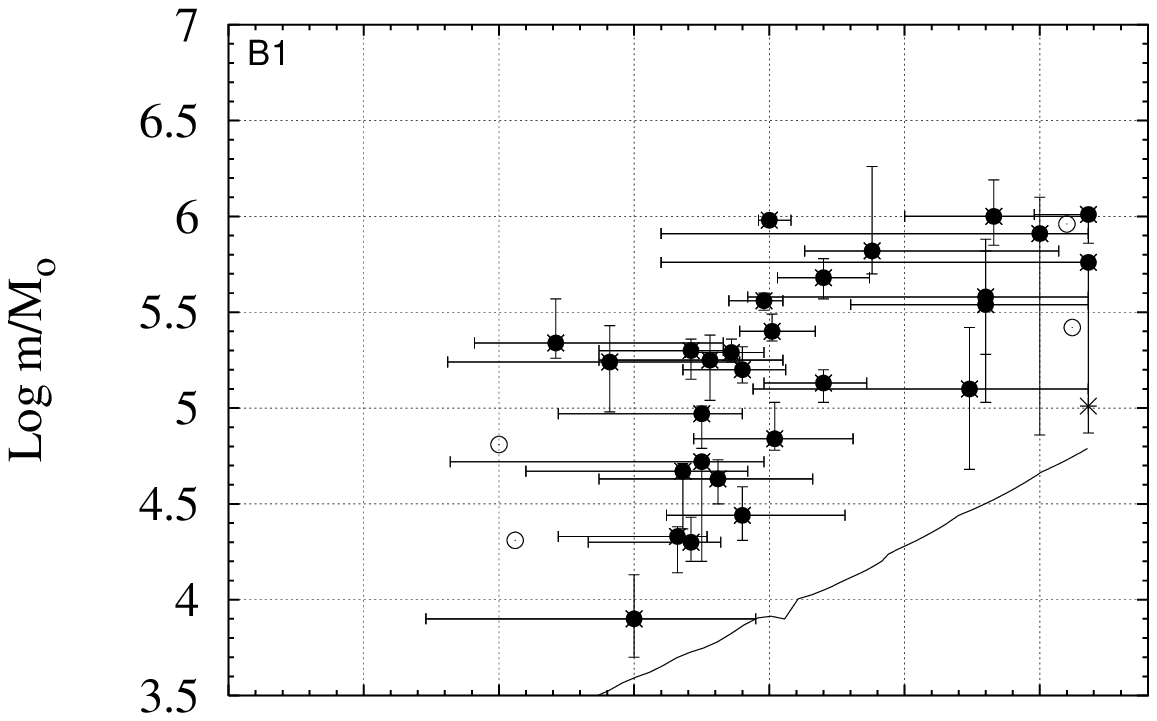, width=\linewidth}
\end{flushright}
\end{minipage}
\vfill \vspace*{-13mm}
\begin{minipage}[b]{\linewidth}
\begin{flushleft}
\epsfig{figure=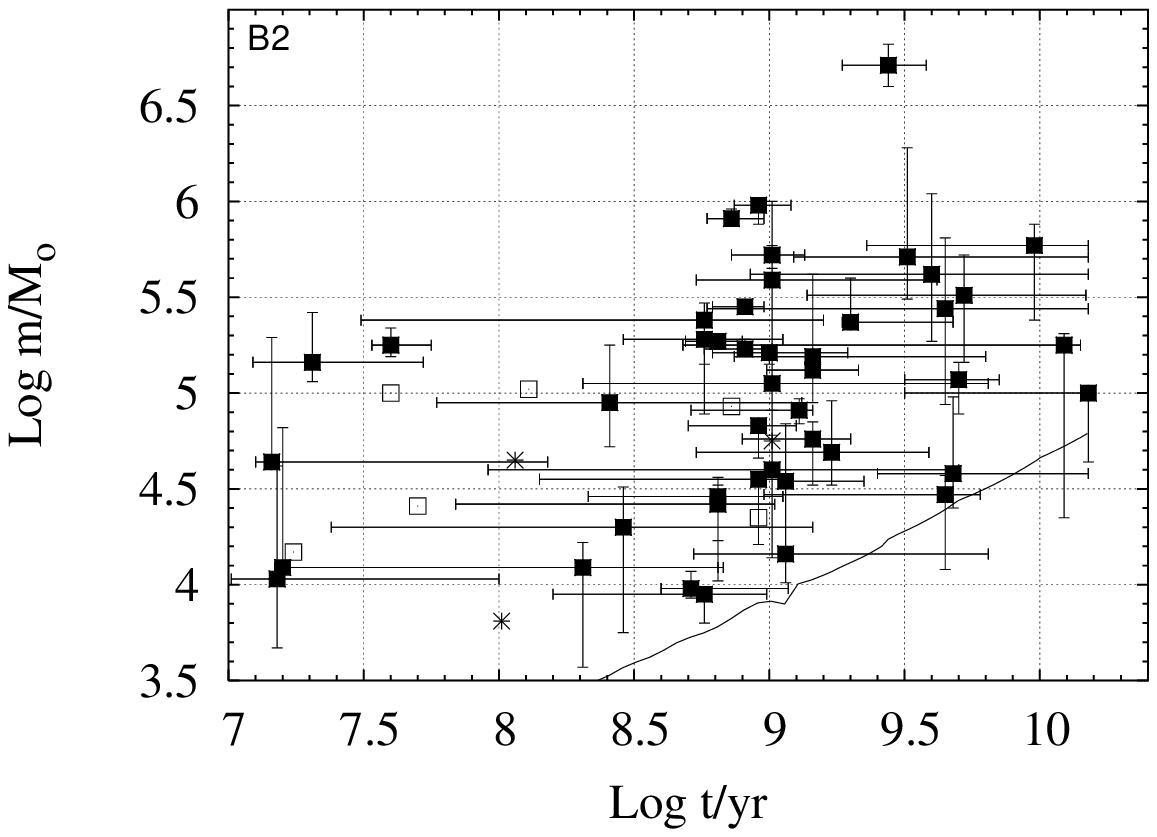, width=\linewidth}
\end{flushleft}
\end{minipage}
\label{fig:LogmLogt}
\caption{Evolution with time of the mass of the detected clusters.
Circles and squares correspond to B1 and B2 clusters respectively.
Open symbols refer to clusters whose $B$, $V$ or $I$ is either
an upper or lower limit.  Asterisks represent clusters with very badly constrained
ages: $t_{\rm min} \simeq 10-20\,$Myr and  
$t_{\rm max} \simeq 10\,$Gyr.  The solid line is the selection limit 
imposed by the detection luminosity threshold for an SSP with 
Z = 0.02 Z$_{\odot}$}  
\end{figure}
Figure 8 shows the evolution with time of the mass of the clusters 
in B1 and B2.  The most striking feature is 
the deficit in low-mass clusters (i.e. ${\rm log}~m/M_{\odot} < 5.5$) 
during the pre-burst phase (log\,$t$/yr $\geq 9.2$) with respect to
the burst period.  However, this effect is likely mainly caused by an  
observational bias, that is, due to the fading of stellar populations 
with time low-mass clusters are more difficult to detect
at old and intermediate age than at young ages (see also 
de Grijs et al.~2002). \\
In order to consider only cluster candidates with good photometry, 
de Grijs et al.~(2001) selected clusters brighter than $V=$~22.5 
(corresponding to the 100\% completeness limit).  The equivalent upper limit in absolute visual
magnitude is $M_V=-$5.3.  The temporal evolution of the corresponding
lower mass limit is shown as the solid line in the two panels of Fig.~8.
This fading line is based on the mass-to-light ratio predicted by the 
BC96 model for a metallicity of Z$\simeq$~0.02\,Z$_{\odot}$.
We can afford to use a model with such a low metallicity, and therefore 
a low mass-to-light ratio, since very
metal-poor clusters are detected at any age up to the burst epoch.
A comparison of the fading line with the distribution of points 
in top and bottom panels of Fig.~8 shows that the deficit in low-mass 
clusters at large age cannot be accounted for solely by the selection 
effect, especially in B1.  

Systems of young clusters are well known to show a cluster initial
mass function scaling as ${\rm d}N \propto m^{-\alpha} {\rm d}m$,
where $m$ is the cluster initial mass and ${\rm d}N$ is the number
of clusters with mass between $m$ and $m + {\rm d}m$.  Observations 
show that the slope $\alpha$ is $\simeq 2$ (Zhang \& Fall 1999;
Whitmore et al. 1999; Bik et al.~2002; see also Sect.~4.6).  
At an intermediate age of 5\,Gyr (log\,$t \simeq 9.7$), the mass detection
threshold is  $3 \times 10^4 {\rm M}_{\odot}$ (log\,$m =4.5$).  If the 
cluster initial mass function had remained unaffected (above the fading 
line), the ratio between the number of clusters with masses 
$3 \times 10^4 {\rm M}_{\odot} \leq {\rm m} \leq 3\times 10^5 
{\rm M}_{\odot} (4.5 \leq {\rm log}\,m/{\rm M}_{\odot} \leq 5.5)$
and the number of clusters with masses     
$3 \times 10^5 {\rm M}_{\odot} \leq m \leq 10^6 
{\rm M}_{\odot} (5.5 \leq {\rm log}\,m/{\rm M}_{\odot} \leq 6)$
would be 
\begin{equation}
\frac{\int_{0.03}^{0.3} m^{-2} {\rm d}m}{\int_{0.3}^1 m^{-2} 
{\rm d}m} \simeq 13\;.
\end{equation}
To make use of the solar metallicity model raises the fading line 
by ${\rm log}\,m \simeq 0.3$ at most in the
${\rm log}\,m$ vs ${\rm log}\,t$ plot through an increase of the
metallicity dependent mass-to-light ratio.  This therefore leads 
to a decrease in the number of detectable clusters whose mass is 
lower than $3\times 10^5\,{\rm M}_{\odot}$.  Even in that case 
however, the proportion between the number of high-mass clusters
($5.5 \leq {\rm log}\,m/{\rm M}_{\odot} \leq 6$) and the number 
of lower-mass clusters (i.e. clusters whose mass is larger than 
the solar metallicity fading line and lower than $3\times 10^5 
{\rm M}_{\odot})$ is still $\simeq 6.5$.  There is an obvious
discrepancy between these theoretical ratios, whatever the metallicity
of the fading line, and the observed ratios as the majority of the clusters
in the pre-burst phase have masses greater than  $3\times 10^5 
{\rm M}_{\odot}$.  We emphasize here that the mass error bars 
do not affect this outcome significantly.  For instance, 
let us consider the old and high-mass cluster, B1-36 
(${\rm log}\,t =10.$, ${\rm log}\,m/{\rm M}_{\odot} = 5.91$; see Table 1).  
The rather large mass error bar suggest that its mass may be as low as 
${\rm log}\,m/{\rm M}_{\odot}=4.86$, so that it may contribute to the low-mass
clusters of the pre-burst phase.  However, its age and its mass errors are
strongly correlated through the age-dependent mass-to-light ratio, that is,
the lower mass limit is coupled with the lower age limit (${\rm log}\,t =8.6$)
and not with the best age estimate (see Sect.~4.2).  As a result, the high-mass 
clusters whose minimal estimate is lower than ${\rm log}\,m/{\rm M}_{\odot}=5.5$
are not expected to contribute to the low-mass clusters class of the
pre-burst phase.    \\

The depletion in low-mass clusters at intermediate age shown by Fig.~8 
is a real effect and illustrates the dynamical evolution of the
M82~B cluster system.  
The disruptive processes working on time-scales on the 
order of a Hubble time include the tidal interactions between the 
clusters and the gravitational field of the parent galaxy and the evaporation 
through two-body relaxation within clusters, the latter being enhanced by
the former.  They affect mostly the low-mass clusters (e.g. Gnedin \& 
Ostriker 1997, de Grijs et al.~2002).
The detection of such evolutionary effects 
is not unexpected because of the dense environment achieved in 
this fossil starburst of the low-mass galaxy M82.
A similar cluster disruption time-scale was derived in the dense centre
of the massive spiral M51 (Boutloukos \& Lamers 2002).  Furthermore, 
de Grijs et al.~(2002) show that the mean densities in M82~B and in the
centre of M51 are remarkably similar, within an order magnitude.
This therefore suggests that cluster disruption time-scales are primarily
driven by the density of the environment in which the clusters are embedded,
irrespective of the mass of the host galaxy.  
Such a result is not unexpected.  The cluster stars which venture to large 
enough distance from the cluster centre are stripped off due
to the tidal influence of the host galaxy, leading to a cluster which is tidally
truncated at a finite tidal radius $r_t$ given by: 
\begin{equation}
r_t = R \left( \frac{m}{3M_{\rm gal}} \right)^{1/3},
\end{equation}     
where R is the galactocentric distance of the cluster and $M_{\rm gal}$ is the mass of the
galaxy within $R$.  Equation (12) also implies that the density of 
a GC, $\rho_{\rm cl}$, cannot be less than three times the average 
density of the host galaxy inside $R$, $\rho_{\rm gal}$, that is 
$\rho_{\rm cl} \geq 3 \rho_{\rm gal}$.  If its orbital motion drives the cluster 
towards higher density regions of the host galaxy, the corresponding shrinkage
of the cluster tidal radius causes a runaway tidal stripping of the outer
cluster stars, thus decreasing the cluster mass and making it even more sensitive
to the galactic tidal field.  Ultimately, this may lead to the cluster's disruption.     

Even in the Magellanic Clouds (MCs), which constitute environments dynamically and tidally 
more gentle than M82~B and M51, do low-mass clusters dissolve with time through normal 
internal process.  Due to the smaller ambient density, the time-scale for depleting low-mass 
clusters is not that short in the MCs however: clusters with masses between 
$10^4$ and $10^5\,M_{\odot}$ survive for a few Gyr (Elson \& Fall 1985b, 1988, Mackey 
\& Gilmore 2002a,b), which is not the case in M82~B.   
In the next section, we show that the 
depletion of low-mass clusters in M82~B is already at work at the burst time.   

%------------------------------------------------
\subsection{Mass functions}
%------------------------------------------------ 

\begin{figure}
\begin{minipage}[b]{0.90\linewidth}
%%\begin{minipage}[b]{0.70\linewidth}
\begin{center}
\epsfig{figure=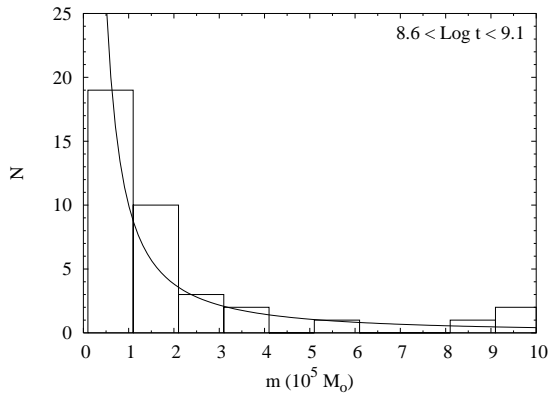, width=\linewidth}
\end{center}
\end{minipage}
\label{fig:MF}  
\caption{Mass histograms of the star clusters (8.5$<{\rm log}~t/{\rm yr}<9.1$) in 
the whole M82 B region and, overplotted, the best-fitting power-law. SSCs whose 
$B$, $V$, $I$ is either an upper or lower limit have been ignored}  
\end{figure}

Figure 9 displays the cluster mass functions for B1 and B2 combined.
We have considered the clusters with ages 8.6 $<$ log\,$t/{\rm yr}$ $<$ 9.1.
To include older clusters in our sample would artificially
bias the cluster mass function towards higher masses because of the
dynamical destruction with time of the low-mass clusters.
De Grijs et al. (2002) derive a characteristic
cluster disruption time-scale for the M82~B cluster system of 
${\rm log}(t_{\rm dis}/{\rm yr}) \simeq 7.5 + 0.62 \times 
{\rm log}(m/10^4\,{\rm M}_{\odot}$).  Obviously, the dynamical 
destruction of star clusters in M82~B takes place on very 
short time-scales.  In fact, the disruption time-scale in M82~B is the 
shortest known in any disc (region of a) galaxy.  Considering 
the clusters formed at ${\rm log}\,t/{\rm yr}=8.6$, the relation between 
the disruption time-scale and the cluster mass mentioned above shows that 
significant disruption must have already affected clusters with masses below 
5.9 $\times$ 10$^5$\,M$_{\odot}$.  For the clusters formed at 
${\rm log}\,t/{\rm yr}$=9.1 (i.e. almost at the onset of the burst),
the same relation indicates that the upper mass of the clusters 
significantly affected by disruption is even 
greater, i.e. 3.8 $\times$ 10$^6$\,M$_{\odot}$.  Thus,  
the dynamical destruction processes have strongly altered
the whole range of the initial mass function of the M82~B clusters
with ages 8.6 $<$ log\,$t/{\rm yr}$ $<$ 9.1.
As a result, the current mass spectrum of these clusters (see Fig.~9) 
no longer represents the formation conditions of the cluster system.  \\

In galaxies hosting systems of young clusters for which deep {\sl HST} 
observations are available, the cluster luminosity functions do not show 
strong evidence for any turnover.  In fact, their shape is consistent, down
to the completeness threshold, with a decreasing power-law of the form 
$\phi (L){\rm d}L \propto L^\alpha {\rm d}L$, where $\phi (L){\rm d}L$
is the number of young star clusters with luminosities between $L$
and $L + {\rm d}L$ and $\alpha$ is the slope of the cluster luminosity 
function.  Since in the age range we consider, 
i.e. $8.6 < {\rm log}\,t/{\rm yr} < 9.1$, the 
mass-to-light ratio does not vary significantly with time and metallicity 
($m/L_V$ $\simeq$ 1), we can directly compare the mass function shown in 
Fig.~9 with the luminosity functions 
derived for other young star cluster systems in merging galaxies or merger
remnants (e.g. NGC~4038/4039, $\phi (L) \propto L^{-1.8}$, Whitmore \& 
Schweizer 1995; NGC~3256, $\phi (L) \propto L^{-1.8}$, Zepf et al.~1999;
NGC~3921, $\phi (L) \propto L^{-2.1}$, Schweizer et al.~1996; NGC~7252,
$\phi (L) \propto L^{-1.8}$, Miller et al.~1997; NGC~3610, 
$\phi (L) \propto L^{-1.9}$, Whitmore et al.~2002).  We have estimated 
the best-fitting power-law exponent with the Levenberg-Marquardt method
(Press et al.~1997).  The mass function in Fig.~9 follows 
${\rm d}N \propto m^{-1.4 \pm 0.2} {\rm d}m$ (see also de Grijs et 
al.~2001).  Our slope is therefore shallower than what has been 
found for many other young star cluster systems 
and shows better agreement with the luminosity function
of the young star cluster population in the Large Magellanic Cloud
($\alpha = -1.5 \pm 0.2$, Elson \& Fall 1985a).  The most likely cause for
such a rather shallow slope is the short disruption time-scale of the
M82~B cluster system.  The clusters first affected by the disruption
are the lowest-mass ones, even though the most massive clusters formed at
the time of the burst have already gone through significant disruption 
processes too.  Due to the shorter disruption time-scale, 
any initial mass function will get biased towards higher masses as time 
goes on and an initial steep slope will turn into a shallower one.
Moreover, the mass distribution of the clusters formed during the burst
shows a turnover at ${\rm log}\,m/M_{\odot} \simeq 5.3$ (de Grijs et al.~2002),
not caused by selection effects (see the location of the fading line 
in Fig.~8).  This also contributes to the difference in slope.

%*************************************************
\section{Conclusions}
%*************************************************
We have estimated the age, the metallicity and the mass of 87 clusters 
located in M82~B, a fossil starburst site in the irregular galaxy M82.
Since this galaxy is located outside the Local Group, our analysis
is based on spectral synthesis models.  We have compared the {\sl HST} 
$BVIJ$ photometry obtained by de Grijs et al.(2001) with the colours of 
simple stellar populations (Bruzual \& Charlot 1996).  
Because the isochrones and the iso-metallicity tracks are not parallel
in a $(V-I)_0$ vs $(V-J)_0$ plot, we have been able to lift  
the age-metallicity degeneracy.     \\
Our results have confirmed the peak in the cluster formation rate 
detected by de Grijs et al.~(2001).  From 1.5 to 0.5\,Gyr ago,
as M82 was getting closer to its large spiral companion M81, M82~B
went through a period of enhanced cluster formation, most probably
induced by the increasing tidal interactions.  We have emphasized its 
chemical evolution showing that M82~B underwent a chemical enrichment 
phase up to super-solar metallicities (i.e. Z$> Z_{\odot}$) about 1 Gyr ago.  
The stellar ejecta from older stellar populations have most probably
enriched the interstellar medium out of which the 1\,Gyr old clusters formed.
At almost the same time however, a subpopulation of very metal-poor 
clusters formed in B2, i.e. the part of M82~B nearest  
the galactic centre.  Their formation may have been stimulated by 
infall, onto B2, of ``fresh'' circumgalactic gas which had escaped the 
chemical enrichment of the interstellar medium while orbiting M82.
The clusters in B1, the eastern part of M82~B, may have been 
self-enriched at the time of their 
formation since they show a mass-metallicity correlation in the sense 
expected by simple self-enrichment models applied to pressure-truncated 
clouds.  In B2 however, a clear-cut relation does not show up.  Due to
the infall of circumgalactic gas, the pressure may not have been stable 
enough to allow the formation of pressure-bound clouds. 
The eastern (B1) and western (B2) parts of M82~B are therefore different 
with respect to both the mass-metallicity diagrams and the formation of
very metal-poor clusters at the time of the burst.
We have also highlighted the dynamical destruction of the 
low-mass clusters over time-scales of the order of several $10^7$yr.  
Finally, we have shown that the mass function of the clusters in a given 
age range obeys a power-law as observed for many cluster systems 
formed in merging galaxies.  However, the slope of the mass function of 
cluster systems associated with mergers is significantly steeper 
($\alpha \simeq -2$) than the slope of the $\simeq$ 1\,Gyr old clusters 
in M82~B ($\alpha \simeq -1.4$).  This is most likely due to 
the disruption processes at work in the M82~B region.  Since the low-mass 
clusters are more quickly destroyed, the mass function will get biased 
towards higher masses and a steep initial mass function will turn
into a shallower one.

\section*{Acknowledgments}
G.P. warmly thanks the Cambridge Institute of Astronomy for its hospitality
where much of this work was carried out.  This research was supported by the
European Community under grant HPMT-CT-2000-00132.  
G.P. is grateful to P.~Magain (Li\`ege Institute) for fruitful advices when tackling
the issue of the error propagation.


\begin{thebibliography}{99}
\bibitem{bik} Bik A., Lamers H.J.G.L.M., Bastian N., Panagia N., Romaniello M.,
Kirshner R., 2002, A\&A, in press (astro-ph/0210594)
\bibitem{brouillet}
Brouillet N., Baudry A., Combes F., Kaufman M., Bash F.,
1991, A\&A, 242, 35
\bibitem{bout} Boutloukos S.G., Lamers H.J.G.L.M., 2002, MNRAS, in press
(astro-ph/0210595) 
\bibitem{bch} Bruzual G., Charlot S., 1996, in: Leitherer C., et al.,
1996, PASP, 108, 996 (AAS CDROM Series 7) (BC96)
\bibitem{card} Cardelli J.A., Clayton G.C., Mathis J.S., 1989, ApJ, 345, 245
\bibitem{dg} de Grijs R., 2001, A\&G, 42, 14
\bibitem{dgOCG} de Grijs R., O'Connel R.W., Gallagher J.S.~III, 2001, 
AJ, 121, 768
\bibitem{dgBL} de Grijs R., Bastian N., Lamers H.J.G.L.M., 2003, MNRAS, 
in press (astro-ph/0211420)
\bibitem{elsonfall} Elson R.A.W., Fall S.M., 1985a, PASP, 97, 692
\bibitem{elsonfallb} Elson R.A.W., Fall S.M., 1985b, ApJ, 299, 211
\bibitem{elsonfall88} Elson R.A.W., Fall S.M., 1988, AJ, 96, 1383
\bibitem{fr} Fall S.M., Rees M.J., 1985, ApJ, 298, 18
\bibitem{freed} Freedman W. et al., 1994, ApJ, 427, 628
\bibitem{gneost} Gnedin O.Y., Ostriker J.P., 1997, ApJ, 474, 223
\bibitem{holtz}
Holtzmann J.A., Watson A.M., Mould J.R., Gallagher J.S. III,
Ballester G.E., Burrows C.J., Clarke J.T., Crisp D., Evans R.W.,
Griffiths R.E., Hester J.J., Hoessel J.G., Scowen P.A.,
Stapelfeldt K.R., Trauger J.T., Westphal J.A., 1996, AJ, 112, 534
\bibitem{kroupa} Kroupa P., Tout C.A.\& Gilmore G.F., 1993, MNRAS, 262, 545
\bibitem{larson} Larson B.L., 1998, MNRAS, 301, 569  
\bibitem{mackgila} Mackey A.D., Gilmore G.F., 2002, MNRAS, accepted for publication
\bibitem{mackgila} Mackey A.D., Gilmore G.F., 2002, in:''New Horizons in Globular Cluster
Astronomy'', ASP Conf.~Series, 2002, eds G.~Piotto, G.~Meylan, G.~Djorgovski, M.~Riello 
\bibitem{marcOc} 
Marcum P., O'Connell R.W., 1996, in From Stars to Galaxies:
The Impact of Stellar Physics on Galaxy Evolution, eds.  C.  Leitherer,
U.  Fritze-von Alvensleben, J.  Huchra (San Francisco: ASP), 419
\bibitem{mclpud} McLaughlin D.E., Pudritz R.E., 1996, ApJ, 469, 194
\bibitem{meylheg} Meylan G., Heggie D.C., 1997, A\&AR, 8, 1
\bibitem{miller} Miller B.W., Whitmore B.C.,
Schweizer F., Fall S.M., 1997, AJ, 114, 2381
\bibitem{ocm} O'Connell R.W., Mangano J.J., 1978, ApJ, 221, 62 
\bibitem{ocghc} O'Connell R.W., Gallagher J.S., Hunter D.A. Colley W.N.,
1995, ApJ, 446, L1
\bibitem{parmj} Parmentier G., Jehin E., Magain P., Neuforge C., Noels A.,
Thoul A., 1999, A\&A, 352, 138
\bibitem{parmgil}
Parmentier G., Gilmore G., 2001, A\&A, 378, 97
\bibitem{press}
Press W.H., Teukolsky S.A., Vetterling W.T. and Flannery B.P. 1997, 
Numerical Recipes (2nd ed.; Cambridge Univ. Press)
\bibitem{prmeyl} Pryor C., Meylan G. 1993, In: S.G. Djorgovski, 
G. Meylan (eds) ASP Conference Series Volume 50, Structure and Dynamics of 
globular clusters, p.~370 
\bibitem{rlrt} Rieke G.H., Loken K., Rieke M.J., Tamblyn P., 1993, ApJ, 412, 99
\bibitem{rosenberg}
Rosenberg A., Saviane I., Piotto G., Aparicio  A., 1999, AJ, 118, 2306
\bibitem{satyap}
Satyapal S., Watson D.M., Pipher J.L., Forrest, W.J. Forrest W.J.,
Greenhouse M.A., Smith H.A., Fischer J., Woodward C.E., 1997, ApJ 483, 148
\bibitem{ssb} Searle L., Sargent W.L.W., Bagnuolo W.G., 1973, ApJ, 179, 427
\bibitem{smwf} Schweizer F., Miller B.W., Whitmore B.C., Fall S.M., 1996,
AJ, 112, 1839
\bibitem{shen} Shen J., Lo K.Y., 1995, ApJ, 445, L99
\bibitem{tinsley} Tinsley B.M., 1972, A\&A, 20, 383
\bibitem{whitschw} Whitmore B.C., Schweizer F., 1995, AJ, 109, 960
\bibitem{whit99} Whitmore B.C., Zhang Q., Leitherer C., Fall M., Schweizer F.,
Miller B.W., 1999, AJ, 118, 1551
\bibitem{wskm} Whitmore B.C., Schweizer F., Kundu A., Miller B.W., 2002,
AJ, 124, 147
\bibitem{wills} Wills K.A., Das M., Pedlar A., Muxlow T.B.W., Robinson T.G., 
2000, MNRAS, 316, 33
\bibitem{worthey} Worthey G., 1994, ApJS, 95, 107
\bibitem{yhl} Yun M.S., Ho P.T.P., Lo K.Y., 1994, Nat., 372, 530
\bibitem{zsa} Zepf S.E., Ashman K.M., English J., Freeman K.C.,
Sharples R.M., 1999, AJ, 118, 752
\bibitem{zf} Zhang Q., Fall S.M., Whitmore B.C., 2001, ApJ, 561, 727
\end{thebibliography}
\end{document}